\documentclass[%
 aip,
rsi,%
 amsmath,amssymb,
preprint,%
]{revtex4-1}

\usepackage{graphicx}
\usepackage{dcolumn}
\usepackage{bm}
\usepackage{xcolor}
\usepackage{tikz}
\usepackage{qcircuit}
\usepackage{braket}
\usetikzlibrary{positioning}
\usepackage{array}
\usepackage{mathtools} 
\usepackage{extarrows}

\usepackage{natbib}

\usepackage{amsmath}
\usepackage{qcircuit}
\begin{document}

\newcommand{\unipd}{Dipartimento di Scienze Chimiche,
 Universit{\`a} degli Studi di Padova, Italy I-35131, Padova, Italy}
\newcommand{\padcen}{Padua Quantum Technologies Research Center,
  Universit{\`a} degli Studi di Padova}

\author{Davide Castaldo}
\email{davide.castaldo@studenti.unipd.it}  
\affiliation{\unipd}
\affiliation{Xanadu, Toronto, ON, M5G2C8, Canada}

\author{Soran Jahangiri}
\affiliation{Xanadu, Toronto, ON, M5G2C8, Canada}

\author{Agostino Migliore}
\affiliation{\unipd}

\author{Juan Miguel Arrazola}
\affiliation{Xanadu, Toronto, ON, M5G2C8, Canada}

\author{Stefano Corni}
\affiliation{\unipd}
\affiliation{\padcen}
\affiliation{Istituto Nanoscienze—CNR, via Campi 213/A, 41125 Modena (Italy)}

\title{A differentiable quantum phase estimation algorithm}




\begin{abstract}


The simulation of electronic properties is a pivotal issue in modern electronic structure theory, driving significant efforts over the past decades to develop protocols for computing energy derivatives. In this work, we address this problem by developing a strategy to integrate the quantum phase estimation algorithm within a fully differentiable framework. This is accomplished by devising a smooth estimator able to tackle arbitrary initial states. We provide analytical expressions to characterize the statistics and algorithmic cost of this estimator. Furthermore, we provide numerical evidence that the estimation accuracy is retained when an arbitrary state is considered and that it exceeds the one of standard majority rule. 
We explicitly use this procedure to estimate chemically relevant quantities, demonstrating our approach through ground-state and triplet excited state geometry optimization with simulations involving up to 19 qubits. This work paves the way for new quantum algorithms that combine interference methods and quantum differentiable programming.
\end{abstract}

\maketitle

\section{Introduction}

Advances in big data information processing \cite{talia2022towards} and sophisticated multiscale modeling \cite{mennucci2019multiscale, ciccotti2022molecular} required by cutting-edge research determine an ever-increasing demand for large-scale computations \cite{chang2023simulations}. This urgent requirement is already driving a major shift in our approach to computation. As the on-chip power dissipation of available semiconductor technologies limits the further development of standard processing architectures, we are likely to see a slowdown of Moore's law. Along the same lines, the diversification of computing paradigms, both from an architectural and algorithmic point of view, could be beneficial for a more sustainable growth in the use of energy and materials\cite{di2023perspective}. 

To some extent, such a transition has already taken place in the context of electronic structure theory. In fact, very active research lines are improving the performance and costs of simulating molecular Hamiltonians by developing implementations on Graphical Processing Units (GPUs)\cite{ufimtsev2008quantum, liu2015quantum, seritan2021terachem} and, more recently, Tensor Processing Units (TPUs)\cite{pederson2022large}. These contributions concern improvements in accuracy and time-to-solution to obtain not only molecular energies but also
other molecular properties which, from the chemist's point of view, can be even more important when it comes to simulating real-life experiments. It is well known that many molecular properties can be expressed in terms of energy derivatives\cite{almlof1985molecular} and can be used for a variety of applications ranging from electronic or magnetic responses to the determination of minimum energy pathways, to the understanding of reaction mechanisms and drug-protein interactions for pharmaceutical purposes and materials discovery\cite{martis2011high, luo2021high}. For these reasons, over the years, much effort has been devoted to developing efficient methods to work out analytical expressions for gradients and higher order derivatives in many fields of quantum chemistry\cite{koch1990coupled, assaraf2000computing, wang2021analytic}.

In the field of quantum computing, the advancement of these methods in a fault-tolerant setting is still lagging behind. Past works aimed at reducing the cost of quantum simulations for ground and excited state energies\cite{hagan2023composite, low2023complexity, babbush2023quantum, kang2022optimized}. Only recently works aimed at extracting physically relevant quantities other than single point energies started to appear\cite{steudtner2023fault}, but very little has been done to study the occurrence of a parametric dependence within the phase estimation circuit.

Here we fill this gap with the development of the first example of an automatically differentiable Quantum Phase Estimation (QPE) algorithm geared to compute energy derivatives for molecular Hamiltonians. Therefore, by merging interference methods and differentiable programming, our contribution unlocks, at once, the quantum simulation of a plethora of molecular properties such as molecular structures or electric and magnetic dipoles. 

Automatic differentiation (AD), a clever combination of the chain rule and dynamic programming, is the workhorse of deep learning techniques\cite{baydin2018automatic} and in recent years it has also proven to be very valuable when applied to quantum chemistry\cite{battaglia2023machine, kasim2022dqc, tamayo2018automatic} and, more generally, to scientific computing\cite{bartlett2006automatic, frank2022automatic, xu2016efficient}. With the advent of Noisy Intermediate Scale Quantum (NISQ) devices, variational algorithms\cite{cerezo2021variational}, in which the circuit is trained to optimize the parameters that define the action of gates tasked with a specific goal, have spurred the development of extensions of automatic differentiation tailored for quantum computing. Much progress has been made to build AD techniques in quantum computing that take into account programs with\cite{zhu2020principles} or without\cite{fang2022differentiable, matteo2023quantum} control flow, leading to the implementation of several differentiable quantum frameworks\cite{bergholm2018pennylane, broughton2020tensorflow}.

Here, we focus on the differentiation through the QPE circuit as described by Nielsen and Chuang's textbook\cite{nielsen2010quantum}, developing a technique for post-processing the measurements that allows for integration into automatically differentiable software. This result is relevant \textit{per se}, as it may provide insights into how to reduce the cost of QPE, which, although asymptotically efficient, is still impractical for demonstrating clear and scalable quantum advantage \cite{lee2023evaluating, fomichev2023initial, rubin2023fault, zini2023quantum}.



This study is organized as follows. Section \ref{intro_idea} is devoted to summarize the QPE algorithm and sketch the proposed differentiable quantum pipeline. In Section \ref{GCE} we discuss the main contribution of this work: building on a previous study of Cruz et al.\,\cite{cruz2020optimizing}, we develop a smooth estimator for the QPE for use with arbitrary input states of the system register. This estimator
allows for seamless integration of the QPE algorithm within a fully differentiable framework. We discuss its properties and derive error bounds to estimate how the smoothing procedure affects the computational cost of eigenvalue and energy gradient evaluation (see Sec.\,\ref{cost_analysis}). Sec.\,\ref{recap_methods} provides a summary of the main options devised so far to compute energy derivatives in a Fault-Tolerant setting with a focus on the results of a recent study of O'brien \textit{et al.}\,\cite{o2022efficient}. In Sec. \ref{results} we apply our method to two example systems, H$_3^+$ and CH$_2$O, performing geometry optimization of their ground singlet and first triplet states. The conclusions
include some remarks on potential roadblocks related to the computational cost of the
differentiable routine and possible ways to circumvent them.

\section{Preliminaries}\label{intro_idea}

We begin with a brief review of the fundamental components of the Quantum Phase Estimation algorithm in its standard formulation\cite{cleve1998quantum}. We will not go into the details of the different simulation techniques used within the QPE circuit as they are beyond the scope of this study. Additionally, the second crucial element of this work involves quantum differentiable programming. Accordingly, we will highlight the main features needed for our development and elucidate why a smooth eigenvalue estimator is crucial in this particular context.

\subsection{Quantum Phase Estimation}\label{intro_to_QPE}

The QPE algorithm\cite{cleve1998quantum} aims to find (some of) the eigenvalues of an arbitrary unitary transformation $U$ with precision $\epsilon$. In this work we will focus on the problem of Hamiltonian simulation, i.e., for us the eigenvalues of $U$ will be directly related to the energies $\{E_i \, | \, i = 1, 2, \dots \, R \}$ of a quantum system encoded in a qubit register (called as system register) of $n$ qubits. This is done by choosing $U$ such that it shares the same eigenstates $\{|i\rangle\}$ with the Hamiltonian $H_{sys}$ of the system and that the eigenvalues of $U$ and $H_{sys}$ are related by an invertible function $\phi_i = f(E_i)$ as shown in Refs.\cite{nielsen2010quantum, cleve1998quantum}.

\begin{figure}[h!]

    \begin{tikzpicture}
        \node at (-12.5, 1) {Readout Reg.};
        \node at (-12.5,-1.35) {System Reg.};
        \draw (0,0) node[anchor=east] {\Qcircuit @C=1em @R=.7em {
            \lstick{\ket{0}} & \qw &  \gate{H} & \ctrl{4} & \qw & \dots &  & \qw & \qw & \multigate{3}{QFT^{-1}} & \meter \\
            \lstick{\ket{0}} & \qw &  \gate{H} & \qw & \ctrl{3} & \dots &  & \qw & \qw & \ghost{QFT^{-1}} & \meter \\
            & \vdots & & & & & \\
            \lstick{\ket{0}} & \qw &  \gate{H} &  \qw & \qw & \dots & & \ctrl{1} & \qw & \ghost{QFT^{-1}} & \meter \\
            \lstick{\ket{0}} & {/}^{\otimes^n} \qw &  \gate{P} & \gate{U(\textbf{x})} & \gate{U^2(\textbf{x})} & \dots & & \gate{U^{2^t}(\textbf{x})} & \qw & \qw & \qw \\
        }};
    \end{tikzpicture}

\caption{Schematic representation of the Quantum Phase Estimation circuit. The readout register comprises $t$ qubits each initialized in the state $|0\rangle$, while the system register is specified by $n$ qubits to which a generic state preparation routine $P$ is applied which sets the state $\psi = \sum_u c_u |u\rangle$. We remark  that the unitary operator $U$ can depend on a set of parameters \textbf{x}. In this study we focus on the task of differentiating the eigenvalues estimated from this circuit with respect to \textbf{x}. }
    \label{qpe_circuit}

\end{figure}
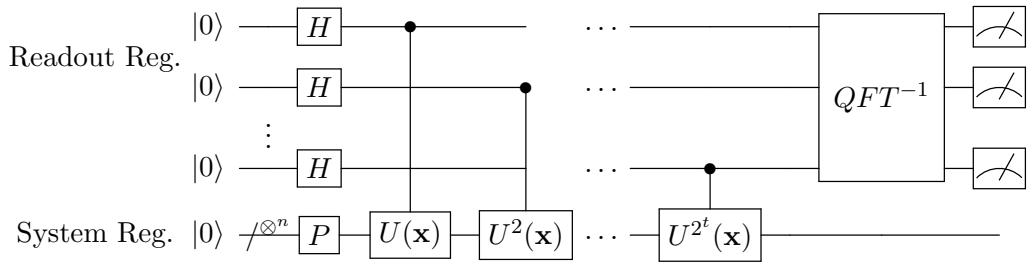

The quantum phase estimation circuit (Fig. \ref{qpe_circuit}) is implemented using two qubit registers: (i) the system register of $n$ qubits encoding the wavefunction of the simulated system and (ii) the readout register of $t$ qubits measured at the end of the circuit execution to estimate phases $\phi_i$.

We can notice from the circuit that the algorithm consists of four main steps. First, the system register is prepared in state $|\psi\rangle = \sum_u c_u |u\rangle$ and the readout register in state $H^{\otimes^t}|0\rangle = \frac{1}{2^{\frac{t}{2}}} \bigotimes_{k = 1}^t (|0_k\rangle + |1_k\rangle) $. After this stage the QPU is in the state

\begin{equation}
    |\Psi\rangle = \frac{1}{2^{\frac{t}{2}}} \sum_u c_u \bigotimes_{k = 1}^t (|0_k\rangle + |1_k\rangle) |u\rangle \, .
\end{equation}
We recall that here the label $u$ runs over the eigenstates of the Hamiltonian and $t$ is the number of qubits in the readout register.

Then the unknown phases are encoded in the quantum circuit by performing a sequence of controlled evolutions of the system register conditioned on the states of the readout register. This brings the QPU into the state

\begin{equation}\label{encoding}
    |\Psi\rangle = \frac{1}{2^{\frac{t}{2}}} \sum_u c_u \bigotimes_{k = 1}^t (|0_k\rangle + e^{i 2 \pi 2^{k-1} \phi_u}|1_k\rangle) |u\rangle \, .
\end{equation}
This state is achieved because the unitary evolution of the system register gives $U |\psi\rangle = \sum_u c_u e^{i 2 \pi \phi_u} |u\rangle$.

After all phases are encoded in the readout register, their values are transferred from the corresponding amplitudes to the computational basis spanned by the readout register by applying an inverse Quantum Fourier Transform\cite{ekert1996shor}. This step is crucial for the whole algorithm. It is the step where the digitalization of the encoded phase occurs. We can think of the computational basis spanned by the readout register as a grid of $2^t$ evenly spaced points that we can use to represent discretely the Fourier transform of the amplitudes encoded in the state of Eq. \ref{encoding}. Now the state of the Quantum Computer (QC) reads:

\begin{equation}\label{after_qft}
    |\Psi\rangle = \sum_u c_u |u\rangle \sum_{b_{1, u} = 0}^1\sum_{b_{2, u} = 0}^1\dots\sum_{b_{t, u} = 0}^1 \big ( \sum_k^{2^t} \frac{e^{i2\pi(\phi_u - 0.\phi_u)(k-1)}}{2^t} \big ) |b_{1, u}\rangle|b_{2, u}\rangle\dots|b_{t, u}\rangle
\end{equation}
In the following, with 0.$\phi_u$ we denote both such binary fraction and the corresponding bitstring $b_{1,u}b_{2,u} \dots b_{t,u}$.

Finally, measuring the state of Eq. \ref{after_qft} expanded on the computational basis, we obtain the parent distribution $ P(0.\phi) = \sum_u P_{\phi_u}(0.\phi) $, where each $P_{\phi_u}$ is the probability of measuring the bitstring $0.\phi$ produced by the eigenphase $\phi_u$:

\begin{equation}
  P_{\phi_u}(0.\phi) =
    \begin{cases}
      1 & \Delta \phi = \phi_u - 0.\phi = 0\\ \frac{g(0.\phi)}{2^{2t}} =
      \frac{1}{2^{2t}}\frac{sin^2(2^t \pi (\phi_u - 0.\phi))}{sin^2(\pi(\phi_u - 0.\phi))} & \Delta \phi = \phi_u - 0.\phi \neq 0\\
    \end{cases}     
\label{probabilities_sampling}
\end{equation}

We will examine in detail the shape of the probability distribution in Sec. \ref{GCE}, as this is the starting point to generalize the circular estimator developed in Ref.\cite{cruz2020optimizing}. At this moment we limit ourselves to highlighting that the result of this type of functional form is to have a superposition of peaks that is narrower as the grid with which we operate the discretization is (e.g., the more readout qubits there are) becomes denser. Each of these peaks corresponds to an eigenstate $|u\rangle$ with non-zero overlap with the initial state $|\psi\rangle$ and is centered on the string that best approximates phase $\phi_u$. As a consequence, the usual strategy for estimating the eigenvalue of the unitary $U$ associated to a given eigenstate $|u\rangle$ is to prepare a state with high overlap with $|u\rangle$ and estimate the corresponding phase $\phi_u$ as $\phi_u \approx \text{argmax}(P(0.\phi)) = 0.\phi_u$. This is often referred to as majority rule estimate. Its accuracy is upper bounded by the number of discretization points used to represent $P(0.\phi)$; in particular the error associated with the estimate is $|\Delta \phi| = |\phi_u - 0.\phi_u| \leq \frac{1}{2^{t+1}}$.

We now move on to discuss quantum gradients and differentiability as they provide us with motivation to go beyond the majority rule estimator in Sec. \ref{GCE}.

\subsection{Differentiating through the Quantum Phase Estimation circuit}

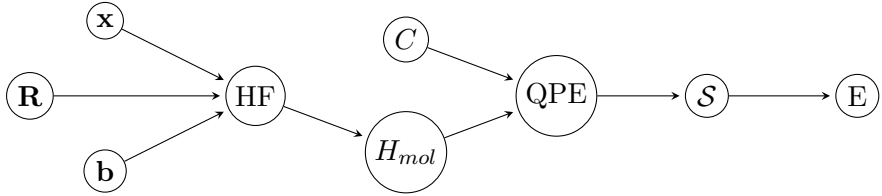
\begin{figure}[h!]

\begin{tikzpicture}[
    every node/.style={draw, circle, inner sep=2pt, minimum size=12pt},
    every path/.style={->, >=stealth, shorten >=2pt}
]

\node (x) at (0,1) {$\textbf{x}$};
\node (y) at (0,-1) {$\textbf{b}$};
\node (z) at (-1,0) {$\textbf{R}$};
\node (f1) at (2,0) {$\text{HF}$};
\node (f2a) at (4,0.75) {$C$};
\node (f2b) at (4,-0.75) {$H_{mol}$};
\node (intermediate) at (6,0) {QPE};
\node (output) at (8,0) {$\mathcal{S}$};
\node (energy) at (10,0) {E};

\draw (x) -- (f1);
\draw (y) -- (f1);
\draw (f1) -- (f2b);
\draw (f2a) -- (intermediate);
\draw (f2b) -- (intermediate);
\draw (z) -- (f1);
\draw (intermediate) -- (output);
\draw (output) -- (energy);

\end{tikzpicture}

\caption{Coarse representation of the hybrid computational graph traversed to compute the molecular energy (and its derivatives) with the QPE algorithm. Each node represents deeply involved functions and, in practice, is written much more accurately. For example, we can consider the quantum circuit of Fig.\ref{qpe_circuit} as a specific representation of the QPE node. Note how, in this representation, the initial state $C$ fed into the QPE circuit is explicitly an independent (differentiable) variable. For an interpretation of this AD pipeline in terms of electronic structure theory quantities, please refer to Appendix\,\ref{physical_meaning}. This work is primarily devoted to developing $\mathcal{S}$ so that it provides a smooth estimator of the molecular energy $E$.}
    \label{computational_graph}

\end{figure}

Differentiable programming is a programming paradigm that leverages the chain rule to obtain the derivative of a computer program with respect to an input parameter. At the basis of this computational paradigm is automatic differentiation, a method for calculating exact numerical derivatives which underlies many machine learning techniques such as the renowned backpropagation algorithm\cite{rumelhart1986learning}. The basic idea behind automatic differentiation is that the computer, regardless of the complexity of the function being computed, executes very simple operations that can all be tracked down and stored in memory. Such a sequence of operations is usually called evaluation trace and can be represented on a computational graph (Fig. \ref{computational_graph}).

The example in the figure illustrates a coarse-grained representation of the computational graph to calculate the energies of molecular electronic states. In Sec.\,\ref{results}, we present results of energy derivation with respect to atomic positions (\textbf{R}), as is required for the optimization of molecular geometries. Here we explicitly set apart the input variables to convey the message that other possible choices lead to natural extensions of the present work. For instance, derivatives with respect to parameters on which the basis set depends (\textbf{b}) or external perturbations (\textbf{x}) are potential avenues to explore. In particular, the latter can enable the calculation of molecular properties relevant to spectroscopy.

A differentiable program is able to compute the derivative of the evaluation trace by traversing the computational graph through all its paths and gathering the partial derivative associated with each node. Specifically, during this operation, referred to as Jacobian accumulation, the program calculates the derivative of the function using the chain rule, which is composed of the derivatives of all the nodes in the computational graph (i.e., the fundamental operations executed by the computer to calculate a function $f$). It is crucial to underline for our purposes that, to maintain the differentiability  of the evaluation trace (and therefore the viability of the algorithm), all individual steps must themselves be differentiable.

We refer to quantum differentiable programming when computational graphs include expressions executed on both classical and quantum hardware. Here, we develop a fully differentiable QPE algorithm and apply it to quantum chemistry, which is one of the most promising fields of application for this algorithm. From this standpoint, the PennyLane software framework represents a natural choice, as it allows for end-to-end differentiability thanks to the differentiable Hartree-Fock solver integrated within its library\cite{arrazola2021differentiable} and allows one to include quantum functions in the computational graph.

In general, quantum functions accept input instructions for the quantum circuit and output results of its execution (i.e., measurements). For our particular purpose we can explicitly write the QPE parent distribution as the result of a parameterized expectation value:

\begin{equation}
\label{qpe_as_fun}
P(0.\phi, \textbf{x}) = \langle 0^{\otimes^t}|\langle \psi| \text{QFT}^{\dagger} \mathcal{U}_{t}^{\dagger}(\textbf{x}) \text{QFT} |0.\phi\rangle \langle 0.\phi| \text{QFT}^{\dagger} \mathcal{U}_{t}(\textbf{x}) \text{QFT} |\psi\rangle |0^{\otimes^t}\rangle
\end{equation}
where $\mathcal{U}_{t}$ is the sequence of controlled evolutions used to encode the phase on the readout register $\mathcal{U}_{t}(\textbf{x}) = \prod_k^t ( |0_k\rangle \langle 0_k| \otimes \mathbf{1}^{\otimes^n} + |1_k\rangle \langle 1_k| \otimes U^{2^k}(\textbf{x}) ) $, and the initial layer of Hadamard gates has been identified as a $QFT$ on the readout register in the product state $|0\rangle^{\otimes^t}$. Further, we used the notation introduced in Eq.\ref{after_qft} to denote the projector on the bitstring state of the readout register corresponding to the binary fraction $0.\phi$.


To perform Jacobian accumulation of hybrid computational graphs it is necessary to develop techniques to compute gradients of quantum functions. Extensive literature focuses on calculating the derivatives of expectation values of parameterized quantum circuits\cite{li2017hybrid, mitarai2018quantum, schuld2019evaluating}. The most relevant for our purposes are the works of Refs.\cite{izmaylov2021analytic, banchi2021measuring, kyriienko2021generalized, wierichs2022general, wiersema2023here}, as they provide general parameter-shift rules to compute derivatives of gates like $U(\theta) = e^{i\theta H}$ with arbitrary $H$. In Sec. \ref{cost_analysis} we will make use of their results to evaluate the number of calls needed to compute gradients by differentiating through a QPE circuit.

Finally, now that we have discussed the fundamentals of differentiable programming, we can note that if we want to build a program that estimates the phase out of the parent distribution induced by a QPE circuit, the classical postprocessing that evaluates an estimator for the phase out of $P(0.\phi)$ must also be differentiable. By definition, the majority rule estimator discussed in the previous section does not satisfy this requirement. In the next section we therefore develop a smooth estimator for QPE.

As anticipated in the introduction, we propose an extension of the work of Cruz et al. \cite{cruz2020optimizing}, producing an estimator we refer to as the Generalized Circular Estimator (GCE).

\section{Generalized circular estimator}\label{GCE}

In this section we present a general smooth estimator for the QPE algorithm, characterize its statistical properties and discuss its application within a differentiable programming framework meant for evaluating derivatives of the molecular Hamiltonian.

Our idea is to combine the Gumbel-Softmax trick\,\cite{jang2017categorical}, used in machine learning to efficiently differentiate across categorical distributions, and the idea of Cruz et al.\,\cite{cruz2020optimizing} of estimating the phase from a circular average\cite{mardia2009directional} of the parent distribution (Eq. \ref{probabilities_sampling}). We begin by recalling the results of Ref.\,\cite{cruz2020optimizing} which are valid when the input state corresponds to an eigenstate $|u\rangle$ of the unitary $U$. 

The authors propose to use the mean phase direction $\mu$ of the first trigonometric moment of the final QPE  QPE measurement distribution as an estimator of the energy. In particular, the first trigonometric moment $\theta$ reads

\begin{equation}
\label{theta_cruz}
    \theta = \sum_{0.\phi}P_{\phi_u}(0.\phi)e^{i 2 \pi 0.\phi} = |\theta| e^{i 2 \pi \mu}
\end{equation}
We recall that $P_{\phi_u}$ is defined by Eq. \ref{probabilities_sampling}.

As pointed out in Ref.\,\cite{cruz2020optimizing}, a nice way to grasp the meaning of $\theta$ is to consider it equivalent to a vector sum in the complex plane, where each vector has a direction specified by $e^{i 2 \pi 0.\phi}$ and a modulus given by $P_{\phi_u}(0.\phi)$. This representation explains another important feature of $\theta$ that also emerges from our generalization, i.e., its accuracy exceeds that of the majority rule estimator. This stems from the fact that each string over- or under-represents the true phase $\phi_u$ because of the discretization in binary fractions, and therefore averaging over many strings weighted by their probabilities allows one to get a more accurate estimate due to error cancellation effects.

Subsequently, the mean phase direction $\mu$ of $\theta$ is given by:

\begin{equation}
    \mu = \text{arg}(\theta)
\end{equation}

Plugging  Eq. \ref{probabilities_sampling} into the definition of $\theta$, the authors get an explicit expression for $\mu$, $\theta$ and $|\theta|$. 

\begin{align}
\label{theta-analytic}
    \theta &= \frac{2^t - 1}{2^t}e^{i2\pi\phi_u} + \frac{1}{2^t}e^{-i(2^t -1)2\pi\phi_u} \\
    \label{mu_original}
    \mu &= \frac{1}{2 \pi} \text{arctan}\Bigg (\frac{(2^t -1) \text{sin}(2\pi\phi_u) - \text{sin}((2^t - 1)2\pi\phi_u)}{(2^t - 1) \text{cos}(2\pi\phi_u) + \text{cos}((2^t - 1)2\pi\phi_u)} \Bigg) \\
    |\theta| &= \sqrt{4^{-t} (4^{t} - 2^{t + 1} + 2 + 2 (2^t - 1)\text{cos}(2^{t + 1} \pi \phi))} 
\label{original_modulus}\end{align}
Both from Eq. \ref{theta-analytic} and Eq. \ref{mu_original} we can see that for large values of $t$ the estimator approaches the true value of $\phi_u$. Further, looking at Eq. \ref{original_modulus} we can see that in the same regime (i.e., large $t$) the modulus of the first trigonometric moment approaches $|\theta| = 1$. Furthermore, this value is approached with exponentially increasing rapidity. As we shall see later, from an algorithmic point of view this behavior means that the additional sampling cost resulting from the definition of this estimator becomes negligible as we increase the number of readout qubits. 

All the above statements about the circular average estimator hold if the input state of our QPE algorithm is an eigenstate of the Hamiltonian. We now extend them to an arbitrary input state.

First of all, it is important to understand why we cannot use Eq. \ref{theta-analytic} directly when $|\psi\rangle$ is not an eigenstate of $U$. As recalled in Sec. \ref{intro_to_QPE}, the probabilities $P(0.\phi)$ result from the contributions of all eigenstates with non-zero overlap with the initial state. 

Thus, if we straightforwardly applied the definition of first trigonometric moment to this new distribution we would obtain:

\begin{equation}
    \theta' = \sum_u |c_u|^2 \sum_{0.\phi}P_{\phi_u}(0.\phi)e^{i 2 \pi 0.\phi} = \sum_u |c_u|^2 \theta_u = \rho' e^{i 2 \pi \mu'}
    \label{arbitrary_state}
\end{equation}
The issue with this expression is that we are now taking a circular average that takes into account the contributions due to different eigenstates. Interestingly, we could show that the last expression leads to an estimator for $\langle \psi |H|\psi\rangle$ as $\langle H \rangle = \text{arg}(\theta')$ . We will not delve into this point, as it is beyond the scope of this study. Nonetheless we point out that comparing the cost of this estimator with existing algorithms for evaluating expectation values in a fault-tolerant setting\,\cite{steudtner2023fault} could be a matter of interest in itself.

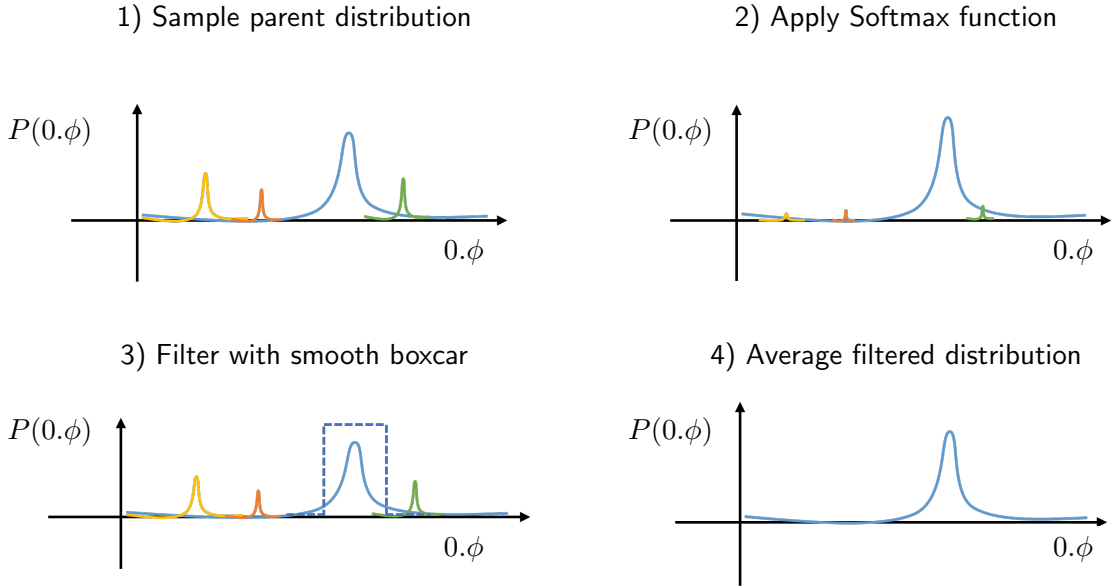
\begin{figure}[h!]

\begin{tikzpicture}[node distance=cm,
    every node/.style={fill=white, font=\sffamily}]

        \node (figure) at (-4,-2) {\centering
    \includegraphics[scale = 0.65]{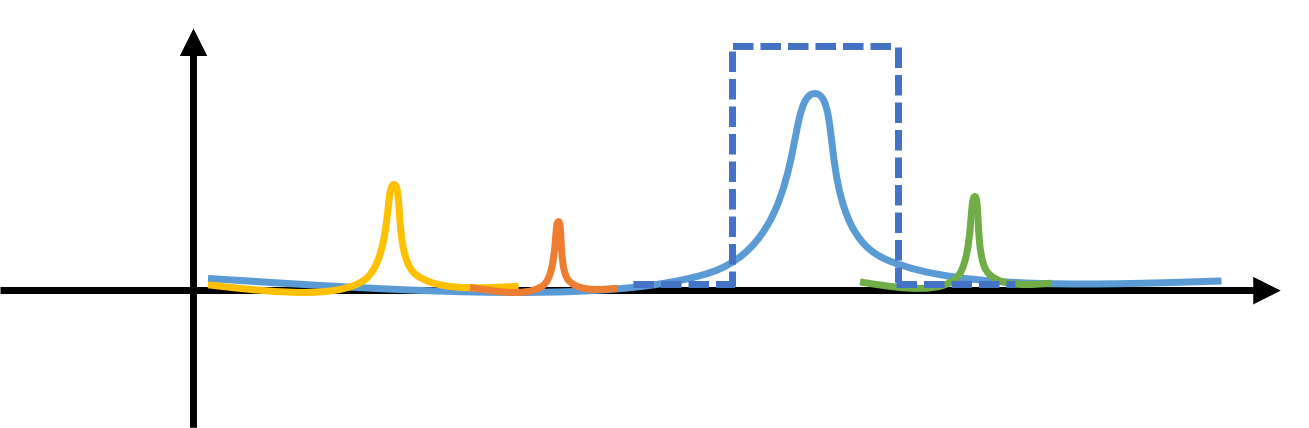}};

        \node (figure) at (4,2) {\centering
    \includegraphics[scale = 0.65]{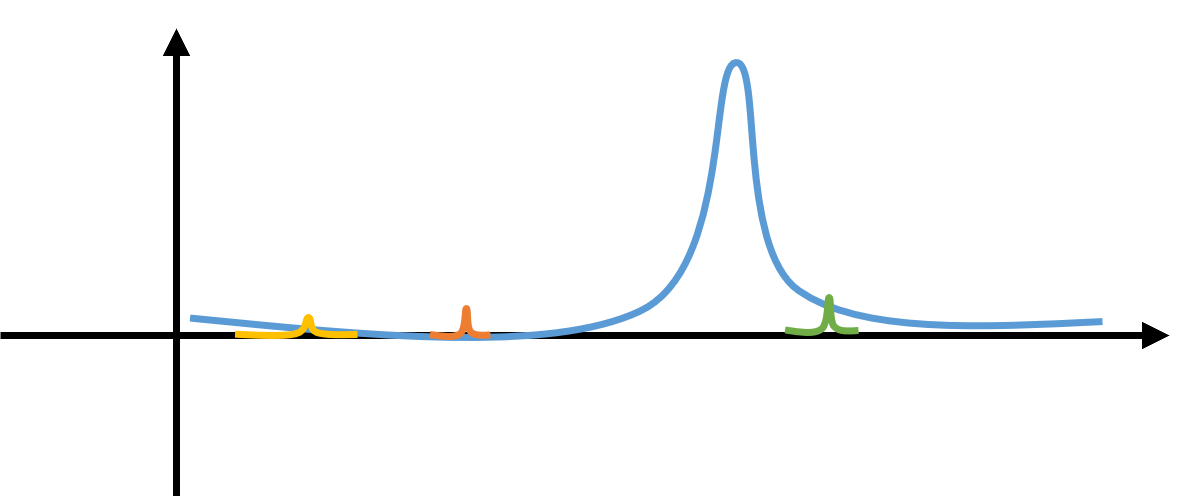}};

        \node (figure) at (-4,2) {\centering
    \includegraphics[scale = 0.65]{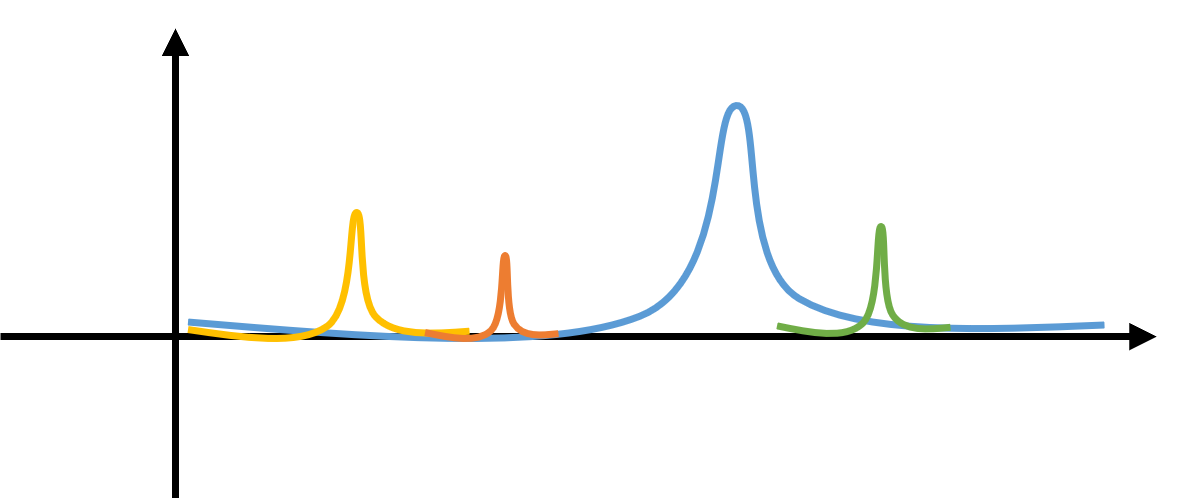}};

        \node (figure) at (4,-2) {\centering
    \includegraphics[scale = 0.65]{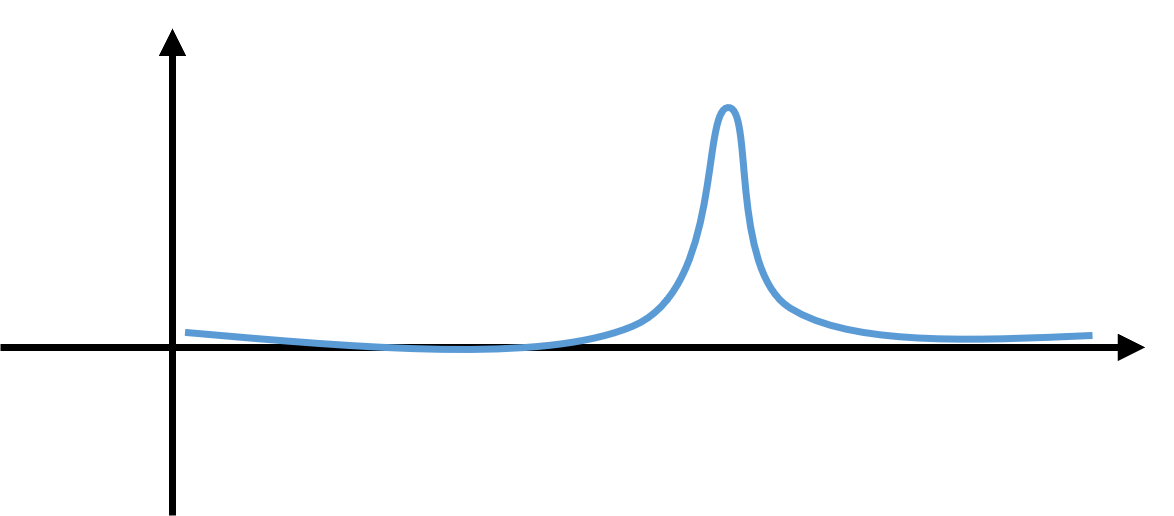}};

    \node (a) at (-4, 4.25) {1) Sample parent distribution};
    
    \node (b) at (4, 4.25) {2) Apply Softmax function};
    
    \node (c) at (-4, -0.25) {3) Filter with smooth boxcar};
    
    \node (d) at (4, -0.25) {4) Average filtered distribution}; 

    \node (f) at  (-7.25, 2.75) {$P(0.\phi)$};

    \node (g) at (-1.75, 1.15) {$0.\phi$};

        \node (h) at  (1, 2.75) {$P(0.\phi)$};

    \node (i) at (6.35, 1.15) {$0.\phi$};

        \node (j) at  (1, -1.25) {$P(0.\phi)$};

    \node (k) at (-1.75, -2.8) {$0.\phi$};

        \node (l) at  (-7.25, -1.25) {$P(0.\phi)$};

    \node (m) at (6.35, -2.8) {$0.\phi$};

\end{tikzpicture}
    \caption{Schematic representation of the postprocessing used to compute the GCE. The lineshapes of the distributions do not reproduce quantitatively Eq. \ref{probabilities_sampling}. (1) Different colors indicate bitstring samplings induced by different eigenstates. Note that, depending on the associated eigenvalues the degree of overlap of two peaks may differ. (2) The effect of the tempered Softmax (Eq. \ref{tempered_softmax}) is to spot the position of the most sampled peak by suppressing all other contributions. (3) Based on the distribution in (2) we are able to build a smooth boxcar function around the peak of interest. (4) Finally, we can estimate the GCE through the convolution of the parent distribution and the boxcar filter.}
    \label{scheme_distributions}
\end{figure}

Eq. \ref{arbitrary_state} is important not only because it provides us with an estimator for an expectation value of a generic $|\psi\rangle$ state, but also because it gives us a clue as to what needs to be done to generalize Eq. \ref{theta_cruz}. Indeed, if we were able to filter all the spurious contributions due to the presence of the states $|u'\rangle \neq |u\rangle$ out of the parent distribution, we could use again the original definition of $\theta$. The strategy we adopt here is as follows: (i) we develop a method to obtain a new distribution similar to the initial distribution except for the contribution of all the $|u'\rangle$s not of interest; (ii) we apply the original definition of $\theta$ to this new distribution to obtain a Generalized Circular Estimator (GCE) $\tilde{\theta}$. These steps are summarized in Fig.\,\ref{scheme_distributions}.

To accomplish the first step we assume that our initial state has the highest overlap $|c_u|^2$ with the eigenstate of interest $|u\rangle$.  This is a standard assumption, since we are usually interested in minimizing the computational cost, which for the majority rule estimator scales as $\mathcal{O}(\frac{1}{|c_u|^2})$.

After that, we can identify the region of the parent distribution induced by $|u\rangle$ using a tempered Softmax function $S_{T}$. This function amplifies differences in the original distribution depending on a parameter $T$ (often referred to as a temperature due to its equivalence to the Boltzmann distribution) according to the following rule:

\begin{equation}
\label{tempered_softmax}
    P'(0.\phi) = S_T(P(0.\phi)) = \frac{e^{-\frac{P(0.\phi)}{T}}}{\sum_{0.\phi} e^{-\frac{P(0.\phi)}{T}}}
\end{equation}

The shape of $P'(0.\phi)$ depends on the chosen temperature $T$ and the initial state $|\psi\rangle$ in the same way as the Boltzmann distribution in statistical mechanics\cite{angeli2013dependence}. In particular, when the temperature is very low we tend to get a one-hot distribution\cite{zheng2018feature} peaked on the most sampled bitstring of the distribution, i.e. $0.\phi_u$. Subsequently, we can extract the $0.\phi_u$ value by weight-averaging the $P'(0.\phi)$ distribution with the binary fraction represented on the computational basis, $0.\phi_u = \sum_{0.\phi} 0.\phi\,P'(0.\phi)$.

\begin{figure}[h!]

\begin{tikzpicture}[node distance=cm,
    every node/.style={fill=white, font=\sffamily}]
    \node (figure) at (0,0) 
    {
    \centering
    \includegraphics[width = \textwidth]{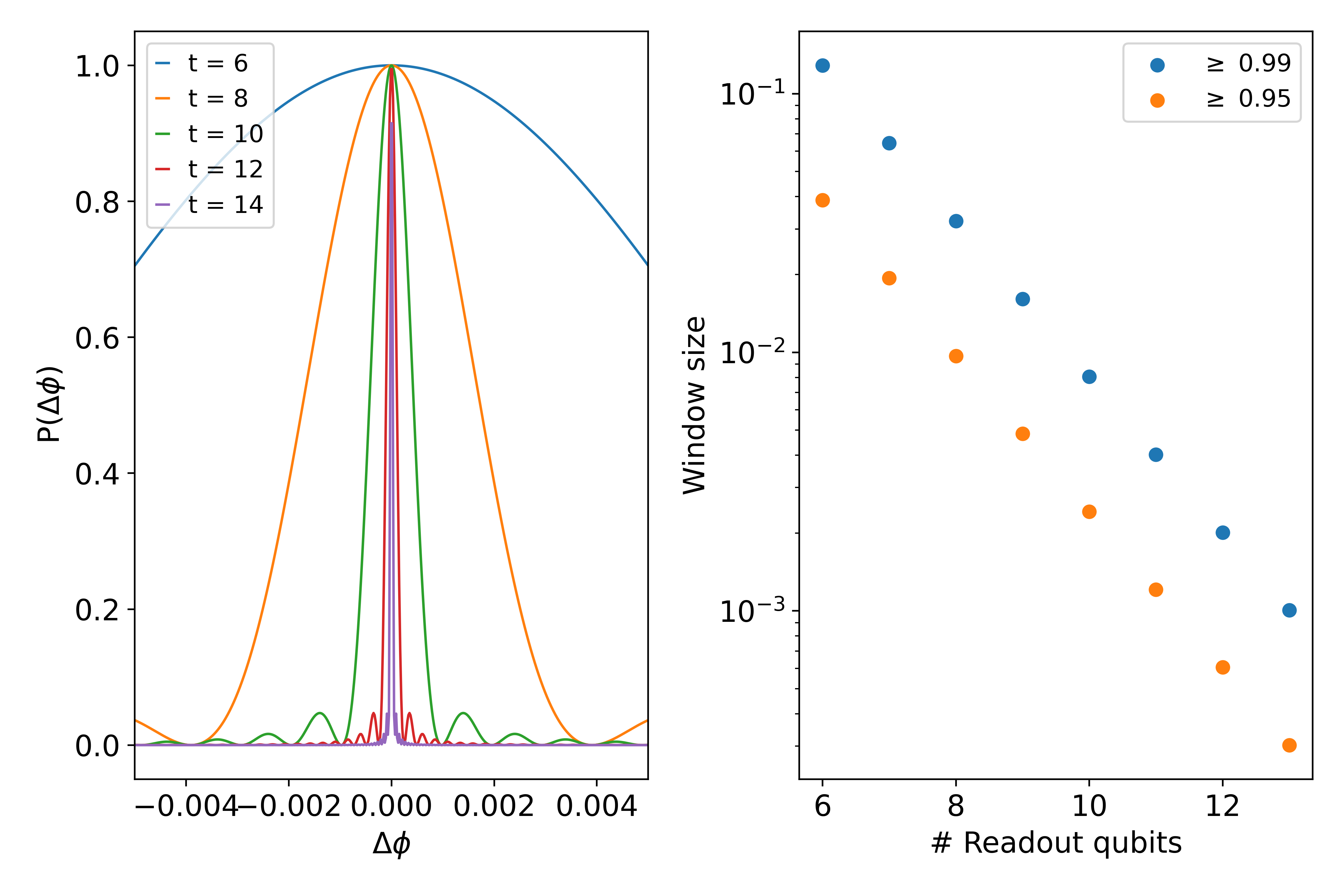}};

\node (a) at (-0.75, 3.65) {(a)};

\node (b) at (7.15, 3.65) {(b)};
    
\end{tikzpicture}
    \caption{a) Plot of the kernel of the quantum phase estimation distribution kernel based on Eq.\,\ref{probabilities_sampling}. The main lobe exponentially shrinks as the number of readout qubits increases. b) Size of the filter window as a function of the number of readout qubits. The values are obtained in such a way that the integrated area of the peak that defines the filter is 95\% (orange dots) and 99\% (blue dots) of the integral over the entire domain, respectively.}
    \label{window_size}
\end{figure}

Once the above is done, we want to identify a subset of bitstrings around $0.\phi_u$ to be used for circular averaging exactly as in the  case of Eq. \ref{theta_cruz}. The reason why we assume we can decouple the contributions of different eigenvalues in populating the same string lies in the definition of $P_{\phi_u}(0.\phi)$. As shown in Fig. \ref{window_size}a, where we plot Eq. \ref{probabilities_sampling} for different values of $t$, $P_{0.\phi_u}$ is a function that shrinks exponentially with the number of readout qubits. The Full Width at Half Maximum ($g_{\text{FWHM}}$) of the central maximum depends on the readout qubits as\cite{moore2021statistical}

\begin{equation}
    g_{\text{FWHM}} = \frac{1}{2^{t}} \, .
\end{equation}

This feature is essential for our filtering procedure as it allows us to identify a subset of binary strings $0.\phi$ for each eigenstate $|u\rangle$ such that the following condition holds: 

\begin{equation}
     P(0.\phi) \approx |c_u|^2 P_{\phi_u}(0.\phi)
     \label{decoupling_approximation}
\end{equation}
The meaning of this statement is that when two eigenphases are far apart the contribution of one of the two to the sampling probability of the given bitstring will be negligible. In other words, given a sufficient number of readout qubits we can always find a bitstring $0.\phi$ for which $P(0.\phi) - \frac{|c_u|^2}{2^{2t}} P_{\phi_u}(0.\phi) < \gamma$, where $\gamma$ is an arbitrarily small number.

These arguments allow us to propose an extension of $\theta$ to arbitrary input states, which we label $\tilde{\theta}$, as:

\begin{equation}
\label{def_generalized_circular_est}
    \tilde{\theta} = |c_u|^2\sum_{0.\phi \in \mathcal{G}}P_{\phi_u}(0.\phi)e^{i 2 \pi 0.\phi} 
\end{equation}
Here we have introduced the set of relevant bitstrings $\mathcal{G} = \{0.\phi_u - h, 0.\phi_u + h\}$, where the width parameter $h$ determines the number of bitstrings included in the circular average.

\begin{figure}[h!]
    \centering
    \includegraphics[width = \textwidth]{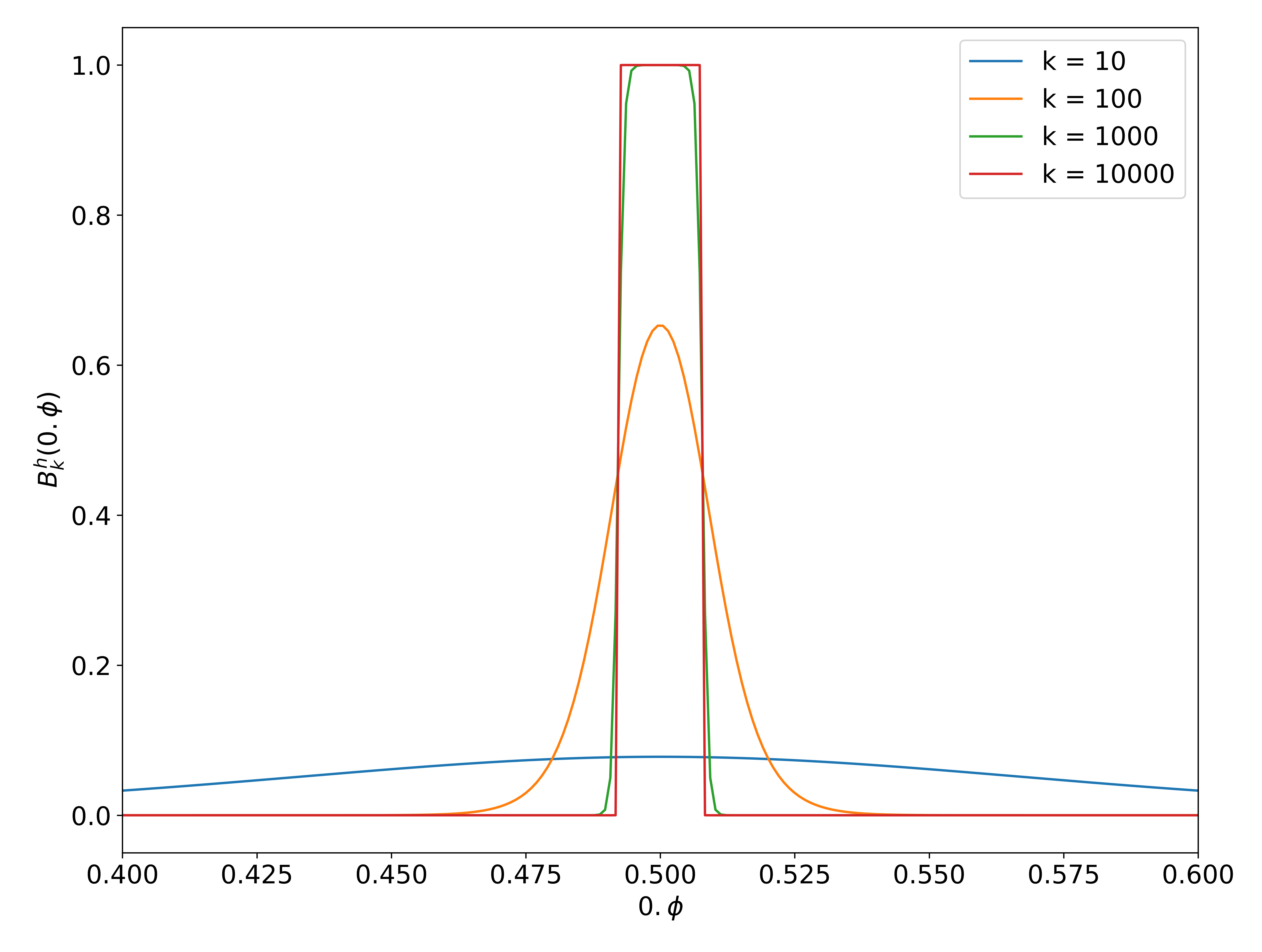}
    \caption{Smooth boxcar filter for different values of k according to Eq.\,\ref{boxcar_filter}. The reported values are obtained for a $t$=10 qubits readout grid with a window width of $|\mathcal{G}|$ = 16 bitstrings corresponding to $h \approx 0.015$.}
    \label{k_effect}
\end{figure}

In practice, when we run the QPE circuit, we can only access measured probabilities. To get a distribution similar to Eq.\ref{def_generalized_circular_est} we apply a smooth boxcar filter centered at $0.\phi_u$:

\begin{equation}
    \tilde{\theta} = \sum_{0.\phi} B_{k}^{h}(0.\phi) P(0.\phi)e^{i 2 \pi 0.\phi}
\end{equation}

where $B_{k}^{h}(0.\phi)$ is given by:

\begin{equation}\label{boxcar_filter}
    B_k^h(0.\phi) = \frac{1}{2} \Big \{ \text{tanh}[k\, (0.\phi - (\phi_{u} - h)] - \text{tanh}[k\,(0.\phi - (0.\phi_{u} + h)] \Big \}
\end{equation}
and $k$ is a tunable parameter which accounts for the steepness of the window function. As we can see (Fig.\ref{k_effect}), the value of $k$ must be chosen appropriately to ensure that the assumptions underlying Eq.\,\ref{def_generalized_circular_est} are valid.

Finally, we obtain the following expressions for the first trigonometric moment $\tilde{\theta}$ and its mean phase direction $\tilde{\mu}$ (which we recall is our estimator for the target eigenvalue):

\begin{equation}
\label{def_generalized_circular_est2}
    \tilde{\theta} = |c_u|^2\sum_{0.\phi \in \mathcal{G}}P_{\phi_u}(0.\phi)e^{i 2 \pi 0.\phi} \approx \sum_{0.\phi} B_{k}^{h}(0.\phi) P(0.\phi)e^{i 2 \pi 0.\phi} = |c_u|^2\rho e^{i 2 \pi \tilde{\mu}}
\end{equation}

\begin{equation}
\tilde{\mu} = \text{arg}(\tilde{\theta}) = \frac{1}{2\pi |\tilde{\theta}|} \text{arctan}(\frac{Im(\tilde{\theta})}{Re(\tilde{\theta})})
\end{equation}

At this point we need to understand how to choose a suitable value for $h$. This should be done by balancing two factors: (i) including as many bitstring as possible, (ii) avoiding spurious contributions from the distributions of other eigenstates'. A possible heuristic to obtain a reasonable value for $h$ is to consider a window size such that the ratio $\frac{I_h(g)}{I(g)}$ between the integrated area of $P(\Delta \phi)$ over the window spanned by $h$ and the integral over the complete domain $I(g)$ is at least 0.95 (see Fig. \ref{window_size}b). Such a strategy may prioritize the first constraint over excluding the distributions of other eigenstates from the circular average. A formal analysis is required to better understand this point.

In the next section we will provide analytical expressions for $\tilde{\mu}$ that will allow us to provide a rationale for this heuristic procedure and demonstrate the accuracy of the generalized circular estimator. Lastly, we will also be able to quantitatively account for the sampling cost associated with the measurement distribution.

\subsection{Statistical analysis of the GCE}\label{gce_formal_analysis}

In this section we derive analytic expressions for the GCE that allow us to highlight its numerical and statistical properties. 

By substituting Eq. \ref{probabilities_sampling} in the definition of $\theta$, in the limit of a large number of readout qubits, the resolution of the grid with which we estimate the phase $\phi$ becomes infinitesimal and we can write

\begin{equation}
\label{continous_passage}
    \tilde{\theta} = \frac{|c_u|^2}{2^t} \int_{0.\phi_u - h}^{0.\phi_u + h} \frac{sin^2(2^t \pi (\phi - x))}{sin^2(\pi(\phi - x))} e^{i 2 \pi x} dx
\end{equation}
the variable $0.\phi$ was changed to the continuous integration variable $x$ and $0 \leq h \leq \frac{1}{2}$.

Compared with the original formulation of Cruz et al., here we only obtain the first term of Eq. \ref{theta-analytic} as we do not carry out the summation but integrate over a continuous variable. The effect of this continuum approximation on the accuracy of our analysis is negligible provided that we consider about $t\geq10$ qubits.

After some algebra we rewrite Eq. \ref{continous_passage} as

\begin{equation}\label{intermediate_passage}
    \tilde{\theta} = \frac{|c_u|^2e^{i2\pi\phi}}{\pi 2^t} \int_{\pi (\Delta \phi - h)}^{\pi (\Delta \phi + h)} \sum_{n, n'= 0}^{2^t-1} e^{2i(n+n'-2^t)u} du 
\end{equation}
where we used Euler equations to manipulate the integral function and changed the integration variable to $u = \pi(\phi -x)$. Furthermore, we introduced $\Delta \phi = \phi - 0.\phi_u$ as a shorthand notation for the mismatch between the real eigenphase and the center of the filtering box.

To solve Eq. \ref{intermediate_passage} we treat separately the terms with $n+n'=2^t$ and $n+n'\neq 2^t$:

\begin{equation}\label{c_solved}
\begin{split}
        \tilde{\theta} &= \frac{|c_u|^2e^{i2\pi\phi}}{\pi 2^t} \Big [(2^t -1) \int_{\pi (\Delta \phi - h)}^{\pi (\Delta \phi + h)} du + \int_{\pi (\Delta \phi - h)}^{\pi (\Delta \phi + h)} \sum_{n, n'= 0 \, , \, n+n' \neq 2^t}^{2^t-1} e^{2i(n+n'-2^t)u} du \Big] \\ 
    & = \frac{|c_u|^2e^{i2\pi\phi}}{\pi 2^t} | \Big [(2^t -1) u - i\sum_{n, n'= 0 \, , \, n+n' \neq 2^t}^{2^t-1} \frac{e^{2i(n+n'-2^t)u}}{2(n+n'-2^t)}  \Big]|_{\pi (\Delta \phi - h)}^{\pi (\Delta \phi + h)} \\
    & = \frac{|c_u|^2e^{i2\pi\phi}}{\pi 2^t} \Big [2\pi(2^t -1)h  - i\sum_{n, n'= 0 \, , \, n+n' \neq 2^t}^{2^t-1} \frac{e^{2i(n+n'-2^t)\pi (\Delta \phi + h)} - e^{2i(n+n'-2^t)\pi (\Delta \phi - h)}}{2(n+n'-2^t)}  \Big] \\
\end{split}
\end{equation}

\begin{figure}[h!]
    \centering
    \includegraphics[width = \textwidth]{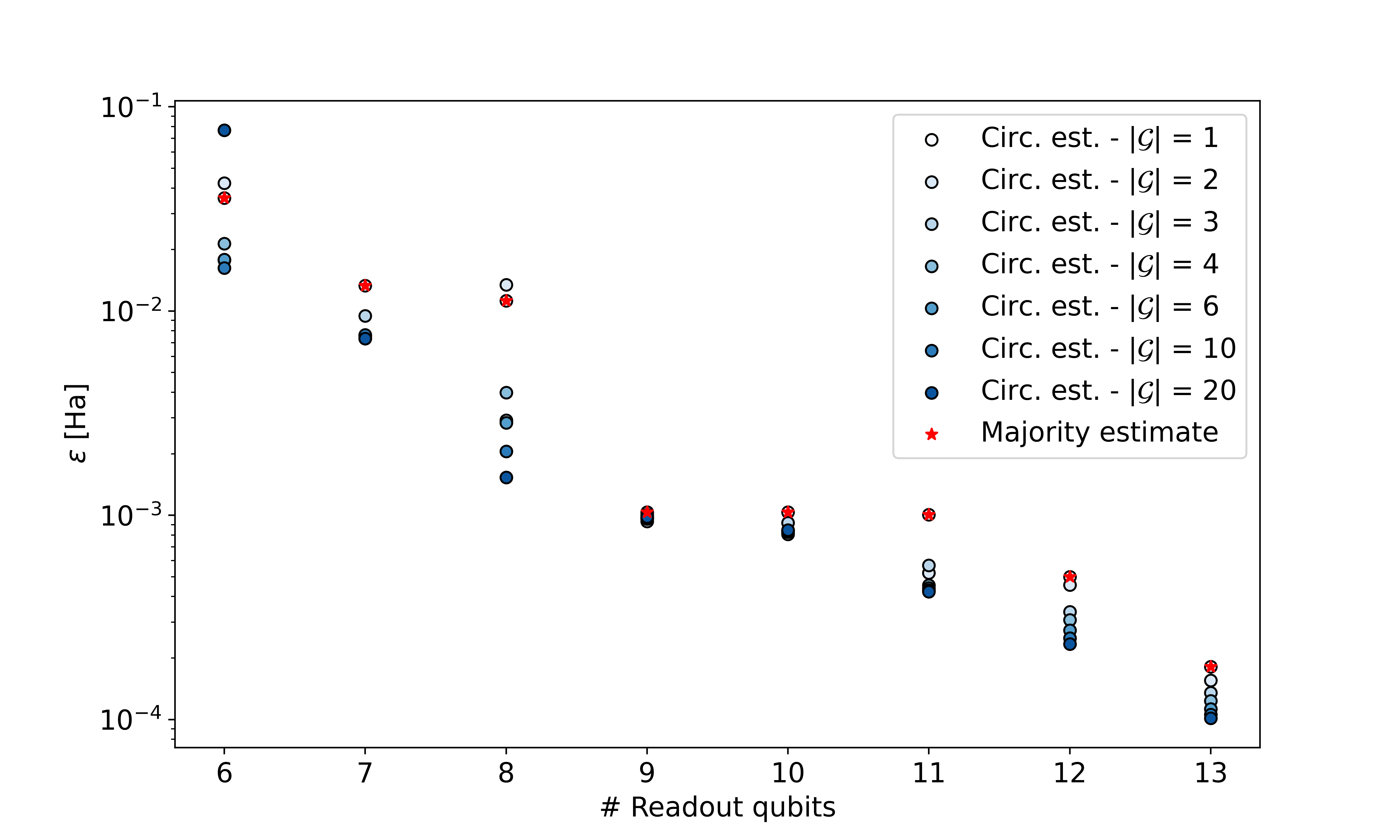}

    \caption{Accuracy of the generalized circular estimator. We report the error $\epsilon$ compared to the FCI solution for the H$_3^+$ molecule in ground state geometry as the number of readout qubits varies. Red stars show the errors obtained using the majority estimator (i.e., considering only the most sampled bitstring), dots display the errors obtained using the estimator proposed here with different window widths (shades of blue).}
\label{window_size_eff}
\end{figure}

We are interested in studying the limiting cases of this formula in which (i) the box spans all possible values of the phase (i.e., $h = \frac{1}{2}$) and (ii) the box is very narrow and of width comparable to the grid spacing (i.e., $h = \frac{|\mathcal{G}|}{2^t}$ where $|\mathcal{G}|$ is the number of bitstrings included in $\mathcal{G}$).  
The first case is interesting because it allows us to recover the original picture of Cruz et al.\cite{cruz2020optimizing}. The second case is useful as it represents the typical regime of use for the GCE.

Starting with the case in which $h=\frac{1}{2}$, it is straightforward to show that $\tilde{\theta} =  \frac{|c_u|^2e^{i2\pi\phi}}{2^t} (2^t -1) $, and therefore

\begin{equation}
\label{large_window}
    |\tilde{\theta}| = |c_u|^2\frac{2^t -1}{2^t} \, \, .
\end{equation}
This is exactly the result of Ref.\cite{cruz2020optimizing} when $|c_u|^2  = 1$ (i.e., the input state is an eigenstate) in the limit of large $t$.

Let us now move to the case in which $h = \frac{|\mathcal{G}|}{2^t}$. We expand the complex exponential functions in the summation up to the second order in close proximity to $\Delta \phi = 0$ and $h = 0$, which holds when the number of readout qubits is reasonably high:

\begin{equation}\label{expanded_theta}
\begin{split}
    \tilde{\theta} & \approx \frac{|c_u|^2e^{i2\pi\phi}}{\pi 2^t} \Big [2\pi(2^t -1)h -\frac{i}{2} ( i4 \pi h \sum_{n, n' = 0}^{2^t -1} 1 -8 \pi^2 \Delta\phi h \sum_{n, n' = 0}^{2^t - 1} (n + n' -2^t) ) \Big ] \\
    & = \frac{|c_u|^2e^{i2\pi\phi}}{\pi 2^t} \Big [2\pi(2^t -1)h + 2 \pi h 2^{2t} - i 4 \pi^2 \Delta\phi h 2^{2t} \Big ] \\ 
    & = \frac{|c_u|^2e^{i2\pi\phi}}{\pi 2^t} \Big [2\pi(2^t -1) \frac{|\mathcal{G}|}{2^t} + 2 \pi |\mathcal{G}| 2^t - i 4 \pi^2 2^t \Delta\phi |\mathcal{G}| \Big ] \\
    & =  \frac{2 |c_u|^2 |\mathcal{G}| |B|}{2^t} e^{i (2 \pi \phi- \alpha)}\\
    & \approx 2 |\mathcal{G}| |c_u|^2e^{i2\pi\phi}
\end{split}
\end{equation}
To pass from the second to the third line we used the fact that $h = \frac{|\mathcal{G}|}{2^t}$. In the fourth line we introduced $B$ as a shorthand for 
\begin{equation}\label{introduced_bias}
     B = |B| e^{-i\alpha} = (1 - i 2 \pi \Delta\phi)2^t + \frac{2^t -1}{2^t} \, \, .
\end{equation}
Finally, the last step in Eq. \ref{expanded_theta} is carried out by considering that $\Delta\phi$ is at most $\frac{1}{2^t}$ and keeping only the leading term in $2^t$. The additional terms originate as an artifact of the smoothing procedure, but it is easy to show that for large $t$ values $|B| = \mathcal{O}(2^t)$ whence $\alpha \equiv \text{arcsin}(\frac{2 \pi}{|B|}) \approx 0$. 

The expression in Eq.\,\ref{expanded_theta}, obtained expanding Eq.\,\ref{c_solved} around small values of $\Delta\phi$ and $h$, usefully shows that the GCE introduces an exponentially vanishing bias into our estimator. In practice, considering the complete expression, the GCE accuracy can exceed that of the standard majority estimator. This result was already found for the case of an exact input eigenstate\cite{cruz2020optimizing}; here we show that it holds for an arbitrary input state provided that the measurement distribution is reconstructed via a sufficient number of samples. For example, in Fig.\,\ref{window_size_eff} we report the error $\epsilon$ made by the GCE in estimating the energy of an electronic state as a function of the number of readout qubits (dots). We compare this error to that made using the standard majority rule (stars). Furthermore, we study the effect of the number of bitstrings included in the circular average, i.e., $|\mathcal{G}|$ (different shades of blue). First of all, we  note that the GCE performs better than the majority rule estimate for all values of $t$. More precisely the error of the latter scales as $\epsilon_{MR} = \frac{1}{2^{t+1}}$ while the circularly averaged estimated error scales as $\epsilon_{GC} = \frac{1}{2^{t+2}}$.

Regarding the dependence of the error on the window size, two effects are noticeable: (i) for relatively small values of $t$, there are some values of $h$ for which we make a larger error than expected using the standard estimator; (ii) contrary to expectations, for low $t$ values it is not true that a wider window necessarily gives a more accurate result. The first effect is related to the point discussed at the end of the last paragraph, that including too many bitstrings when peaks corresponding to different eigenstates are not sufficiently separated can introduce spurious errors. The second effect stems from the fact that the improved accuracy of the GCE is due to error cancellation effects. With a lower number of qubits, this error cancellation may undergo significant variation due to discretization effects on the averaged distribution $P(0.\phi)$.


We now conclude this section by answering the following question: how many samples are necessary to reconstruct the parent distribution with the accuracy shown in Fig.\,\ref{window_size_eff}?

\begin{figure}[h!]

\begin{tikzpicture}[node distance=cm,
    every node/.style={fill=white, font=\sffamily}]

    
    \node (figure) at (0,0) 
    {\centering
    \includegraphics[width = \textwidth]{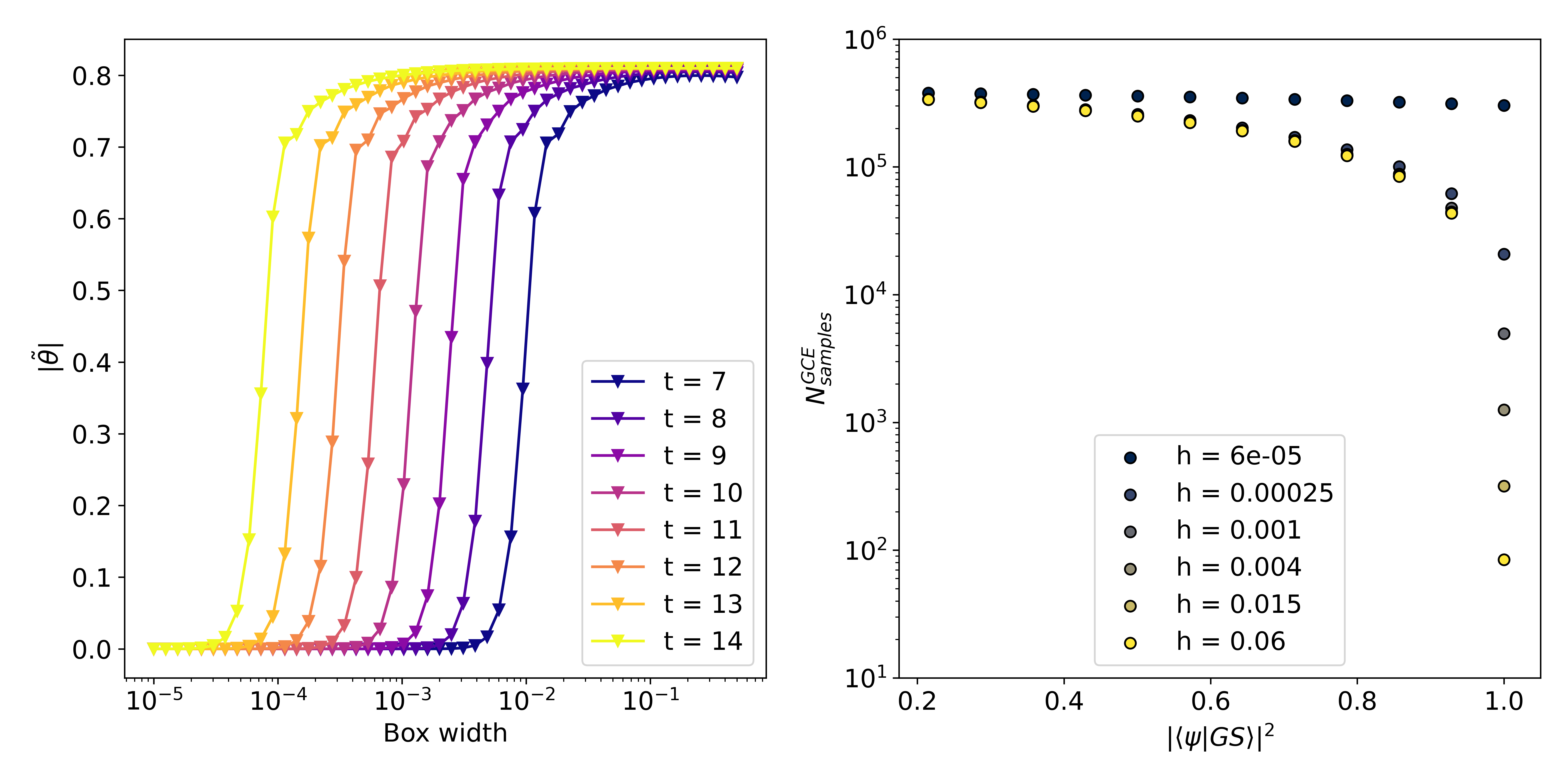}};

\node (a) at (-6.5, 2.45) {(a)};

\node (b) at (1.75, 2.45) {(b)};

\end{tikzpicture}
    \caption{Sampling cost of the generalized circular estimator. a) Modulus of $\tilde{\theta}$ as a function of the box width for different numbers of readout qubits according to Eq.\,\ref{c_solved}. We set $\Delta \phi = \frac{1}{2^t}$. We also consider an initial state with an overlap with the target state of $|c_u|^2 = 0.8$. b) Sampling cost of the generalized circular estimator $N_{samples}^{GCE}$ as a function of the overlap of the initial state with a target ground state $|\langle\psi | GS \rangle|^2$ for varying window width $h$.}
\label{optimal_window_size}
\end{figure}

To this end, we compute the variance of the mean phase direction $\tilde{\mu}$ and use Chebyshev's inequality:

\begin{equation}
\label{mu_samples}
    N^{GCE}_{samples} \geq \frac{\text{Var}(\tilde{\mu})}{\epsilon^2}
\end{equation}

The variance of the mean phase direction is obtained by applying the error propagation rule to $\tilde{\theta}$ (Eq.\,\ref{variance_mu}), assuming the latter is a proper random variable\,\cite{JanComplexVariable} and using the definition of variance from circular statistics\,\cite{mardia2009directional} (Eq.\,\ref{variance_theta}):

\begin{equation}
\label{variance_mu}
    \text{Var}(\tilde{\mu}) = \frac{1}{2} \Bigg [ |\frac{\partial \tilde{\mu}}{\partial Re\{ \tilde{\theta} \}}|^2 \text{Var}(\tilde{\theta}) + |\frac{\partial \tilde{\mu}}{\partial Im\{ \tilde{\theta} \}}|^2 \text{Var}(\tilde{\theta}) \Bigg ] 
\end{equation}

\begin{equation}
\label{variance_theta}
        \text{Var}(\tilde{\theta}) = 1 - |\tilde{\theta}|
\end{equation}

In Fig.\,\ref{optimal_window_size} we study the cost of obtaining an estimate for the mean phase direction with an error of $\epsilon = 1\,mHa$. In particular, in Fig.\,\ref{optimal_window_size}a we plot $|\tilde{\theta}|$ with varying number of readout qubits and box width $h$. We can see that the effect of increasing the number of readout qubits is to recover the complete information of the peak (that is, the limit value of the quantity in Eq.\,\ref{large_window} is reached) with a smaller window. As can be seen from Eqs.\,\ref{mu_samples}-\ref{variance_theta}, this directly impacts the cost of the routine. Indeed, as we approach the situation of Eq.\ref{large_window}, we tend to lower the estimation cost, and to do so carefully (i.e., without including contributions of other peaks in the average) we need to increase the grid resolution. Similar considerations can be drawn from  Fig.\ref{optimal_window_size}b where we see the effect of the initial state overlap on the cost while varying the window width at fixed number of readout qubits ($t=13$). Numerical analysis is presented in appendix\,\ref{num_sampling_noise}.

Overall we see that the increased accuracy of $\tilde{\mu}$ has a considerable computational cost, especially if the input state has a small overlap with the eigenstate of interest. Further investigation is required to reduce such a computational overhead and thus enhance the scalability of the approach with respect to the size of the system. The improvement would be particularly relevant to the treatment of strongly correlated systems, where, even for modest system size, the initial state is generally expected to have a relatively small overlap with the target state\cite{fomichev2023initial}. Finally, we recall that beyond an improved accuracy this approach allows to obtain gradients of the estimated eigenvalues. This is a completely different task and in the next section we go through the costing of this routine and compare with existing methods in literature.

\section{Cost of gradient estimation}\label{cost_analysis}

Building on the previous analysis, in this section we analyze the cost of evaluating the gradient with the quantum differentiable program sketched in Fig.\,\ref{computational_graph} compared to state-of-the-art methods for evaluating energy gradients for molecular Hamiltonians on a fault-tolerant quantum computer.

\subsection{Computational cost analysis}

The overall cost for the gradient evaluation with the differentiable QPE is given by:

\begin{equation}
    \Tilde{C} = CNk
\end{equation}
where $C$ is the cost of a single QPE circuit evaluation, $N$ is the number of calls to the routine needed to evaluate $\tilde{\mu}$ with an $\epsilon$-good estimate and $k$ is the number of calls required for the gradient evaluation. In this section we estimate $\Tilde{C}$ breaking down the different contributions. 

Regarding molecular systems, we note that the scaling of the computational cost of the routine with system size (meant as the number of electrons or spin-orbitals) is included in the quantity $C$ and is not considered here. As for $N$, by definition we have $N=N^{GCE}_{samples}$ as given by Eq.\,\ref{mu_samples}.

Thus, all that remains is to estimate $k$. The number of calls to the QPE subroutine to estimate the gradient with a parameter-shift rule, following Wierichs \textit{et al.}\cite{wierichs2022general}, is given by $k = RM$ where $M$ is the number of Cartesian coordinates  (or more generally the number of variables) with respect to which the differentiation is performed and $R$ is a factor counting either the number of parameter-dependent gates (decomposition based parameter-shift rule) or the number of unique positive differences in the parameterized circuit spectrum (Fourier based parameter-shift rule). Depending on the structure of the Hamiltonian the number of independent circuit executions required by the parameter-shift rule can vary substantially. Since the eigenvalue distribution for a molecular Hamiltonian does not have an ordered structure, in the cost analysis it is safer to consider in the cost analysis a decomposition based strategy for the quantum gradient. If we denote $G$ the number of gates where the parameters appear after the compilation of the simulation routine implemented in the quantum circuit, we obtain: 

\begin{equation}
    k \approx \mathcal{O}( G \, t \, M)
\end{equation}
where $t$ is the number of readout qubits and therefore of calls to the quantum simulation routine within the phase estimation circuit.

The last step needed to estimate the overall cost $\tilde{C}$ is to obtain the number of shots required for the gradient estimation. As we focus on the quantum chemistry framework, we label $\textbf{R}$ (to indicate nuclei positions) the parameters with respect to which we take the gradient. Incidentally, it is worth emphasizing once again that the scope of application of energy derivatives and automatic differentiation for quantum chemistry is much broader than just geometry optimization. 

Within the (either decomposition-based or Fourier-based) parameter-shift framework we can write the force acting on the nucleus at the position $R_j$ using the generalized circular estimator as:

\begin{equation}
    \frac{\partial \tilde{\mu}}{\partial R_j} = \sum_l^k y_l \tilde{\mu}(\textbf{R}_l)
\end{equation}
where $k$ is the number of function evaluation required by the parameter-shift rule and $\textbf{y}$ is a coefficient vector defined by the particular flavour of gradient rule used.

The optimal budget shot allocation to obtain $\frac{\partial \tilde{\mu}}{\partial R_j}$ within a tolerable error $\epsilon$ is:

\begin{equation}
    N_{\partial_{j} \tilde{\mu}} = \frac{\text{Var}(\tilde{\mu}) || \textbf{y}||^2_1}{\epsilon^2}     \label{shots-gradient}
\end{equation}

Finally, the number of queries to a circuit that implements the evolution operator $U$ for gradient estimation with a fully differentiable Quantum Phase Estimation can be written as:

\begin{equation}
    \Tilde{C} = \mathcal{O}( G \, t \, M N_{\partial_{j} \tilde{\mu}})
    \label{cost-gradient}
\end{equation}

Interestingly, lowering the number of readout qubits affects simultaneously: (i) $t$, (ii) $\epsilon$ and (iii) $N_{\partial_{j} \tilde{\mu}}$ as we expect that the 1-norm of the parameter-shift rule's coefficients decreases as their number decreases. Methodologies such as those developed in Ref.\cite{rendon2022effects} may be extremely useful in this context.



\subsection{Related literature on gradient estimation}\label{recap_methods}

The goal of this section is to summarize previous related studies on the evaluation of energy derivatives within the context of quantum computation. We refer to the study of O' Brien \textit{et al.}\cite{o2022efficient} for a more detailed report on state-of-the-art methods to compute energy gradients in a fault-tolerant setting.

We remark that all our considerations are made for a fully variational wavefunction in the sense that, given a set of \textit{external physical parameters} \textbf{x} and a set of \textit{wave function parameters} $\lambda$ we calculate the energy from the expression 

\begin{equation}
    E(\textbf{x}) = E(\textbf{x}, \lambda^*)
\end{equation}
where the parameter set $\lambda^*$ represents the optimal value of $\lambda$ and where the optimized energy function $E(\textbf{x}, \lambda^*)$ satisfies the variational conditions for all values of the external parameters $\textbf{x}$. This statement also holds for the quantities extracted from QPE since, within the discretization error due to the finite number of readout qubits, we sample from the exact eigenbasis of the molecular Hamiltonian.

\subsubsection{Hellmann-Feynman approach}\label{hellman-feynman}

This first approach is based on the Hellmann-Feynman theorem\,\cite{feynman1939forces}, which expresses the derivative of the total energy with respect to a parameter (a nuclear coordinate in this case) as the expectation value of the derivative of the Hamiltonian operator with respect to that same parameter:

\begin{equation}
\label{hf-theorem}
    \frac{dE}{d\textbf{x}} = \langle \Psi | \frac{dH}{d\textbf{x}} | \Psi \rangle 
\end{equation}

Eq. \ref{hf-theorem} does not provide a recipe for directly computing energy gradients but only tells us the quantity that we need to calculate. Among the various strategies proposed in the literature, a possible strategy is to use the overlap estimation algorithm which allows one to compute Eq. \ref{hf-theorem} at the Heisenberg limit (i.e., with a cost scaling as $\mathcal{O}(\epsilon^{-1})$). This method provides an estimate of the energy gradient by directly calculating the expectation value with the algorithm proposed in Ref.\,\cite{knill2007optimal}. However, since we are dealing with a real expectation value we only need to call the amplitude estimation algorithm twice instead of three as originally sketched in Ref.\cite{knill2007optimal}.

It is important to notice that this method requires using a block-encoding routine for the energy derivative operator. Using tensor hyper-contraction the authors of Ref.\,\cite{o2022efficient} were able to show that, despite a larger prefactor, the asymptotic cost of block-encoding $\frac{d H}{d \textbf{x}}$ equals the asymptotic cost of block-encoding the molecular Hamiltonian.

\subsubsection{Direct gradient-based quantum estimation algorithm}\label{qsvt-gradient}

Another ad hoc approach developed for computing gradients of functions \cite{o2022efficient} can be used as an alternative to the previous method. Building upon the algorithm developed in \cite{gilyen2019optimizing} and refined in \cite{huggins2022nearly}, the authors of \cite{o2022efficient} demonstrate that nearly optimal scaling can be achieved for calculating the forces of molecular Hamiltonians.

Without going into details the main idea is that using a single query to a state preparation routine $U_{\psi}$ (which implies that we know how to prepare the exact ground state of the molecule) and querying $\mathcal{O}(\frac{\sqrt{M}}{\epsilon})$ (where M is the number of variables to differentiate) times the oracle 

\begin{equation}
    U_p : |x\rangle|0\rangle \rightarrow |x\rangle \otimes (\sqrt{f(x)} |\psi_1(x)\rangle |1\rangle + \sqrt{1-f(x)} |\psi_0(x)\rangle |0\rangle)
\end{equation}
it is possible to simultaneously calculate a given $f$(x) at different points and compute a finite difference approximation of the function directly on the quantum computer.

This algorithm, which uses different qubit registers, makes several calls to routines that implement operators like $U = e^{-i t \frac{dH}{dR_k}}$, to which the considerations in the previous paragraph on block-encoding apply.

\subsubsection{Finite difference approach}


 Finally, another possibility is to consider the QPE algorithm itself as a black-box which the function to be sampled at a given point in the parameter space. From this standpoint, we can use this algorithm to compute gradients via classical finite difference approximations.

Following Ref. \cite{o2022efficient}, we write the finite difference formula of degree 2$m$ as:

\begin{equation}
\begin{split}
\label{finite-diff}
        \frac{dE^{(m)}}{d\textbf{x}} & = \sum_{l = -m}^{l = m} b_{l}^{(m)} E(\textbf{x} + l\,\delta\textbf{x} \cdot \textbf{v}) \\
     b_l^{(m)} & =
    \begin{cases}
      0 & l = 0\\  \frac{(-1)^{l-1}}{\delta\textbf{x} \, l} \frac{\binom{m}{|l|}}{\binom{m+|l|}{|l|}} & l \neq 0\\
    \end{cases}     
\end{split}
\end{equation}

While details on the computational cost of this algorithm can be found in Ref.\,\cite{o2022efficient}, here we want to highlight the differences that arise from tackling Eq.\,\ref{finite-diff} with the ME and GCE. 
Applying error propagation to the expression for the energy derivative in Eq.\,\ref{finite-diff} we obtain:

\begin{equation}
    \Big | \frac{dE}{d\textbf{x}} -  \frac{dE^{(m)}}{d\textbf{x}} \Big | = \sum_{l = -m , l \neq 0}^{l = m} \epsilon_{PE} |b_l|^2 + \epsilon_{FD}
\end{equation}
Two sources of error appear in the last equation: (i) the error due to the phase estimation procedure $\epsilon_{PE}$ and (ii) the error due to the finite difference approximation $\epsilon_{FD}$. At first, we focus on the error due to the phase estimation procedure. When using the ME estimator, $\epsilon_{PE} = \epsilon_{ME} = \frac{1}{2^{t+1}}$. Using the GC estimator instead, it is $\epsilon_{PE} = \epsilon_{GC} = \frac{1}{2^{t+2}} + \sqrt{\frac{\text{Var}^{h, t}(\mu)}{N}}$. From the last two error expressions, we deduce that the gradient approximation obtained with the GC estimator is more accurate if

\begin{equation}
    \sqrt{\frac{\text{Var}^{h, t}(\mu)}{N}} \leq \frac{1}{2^{t+2}} \, \, .
\end{equation}

At this point it is important to emphasize that, when using the ME, the step size $\delta\textbf{x}$ used must satisfy

\begin{equation}
    |\phi(\textbf{x} + l\,\delta\textbf{x} \cdot \textbf{v}) - \phi(\textbf{x} + l'\,\delta\textbf{x} \cdot \textbf{v})| > \frac{1}{2^t} \, \, \forall \, l, l'
\end{equation}

\begin{figure}[h!]
   
\begin{tikzpicture}[node distance=cm,
    every node/.style={fill=white, font=\sffamily}]
    
    \node (figure) at (0,0) 
    {\centering
    \includegraphics[width =\textwidth]{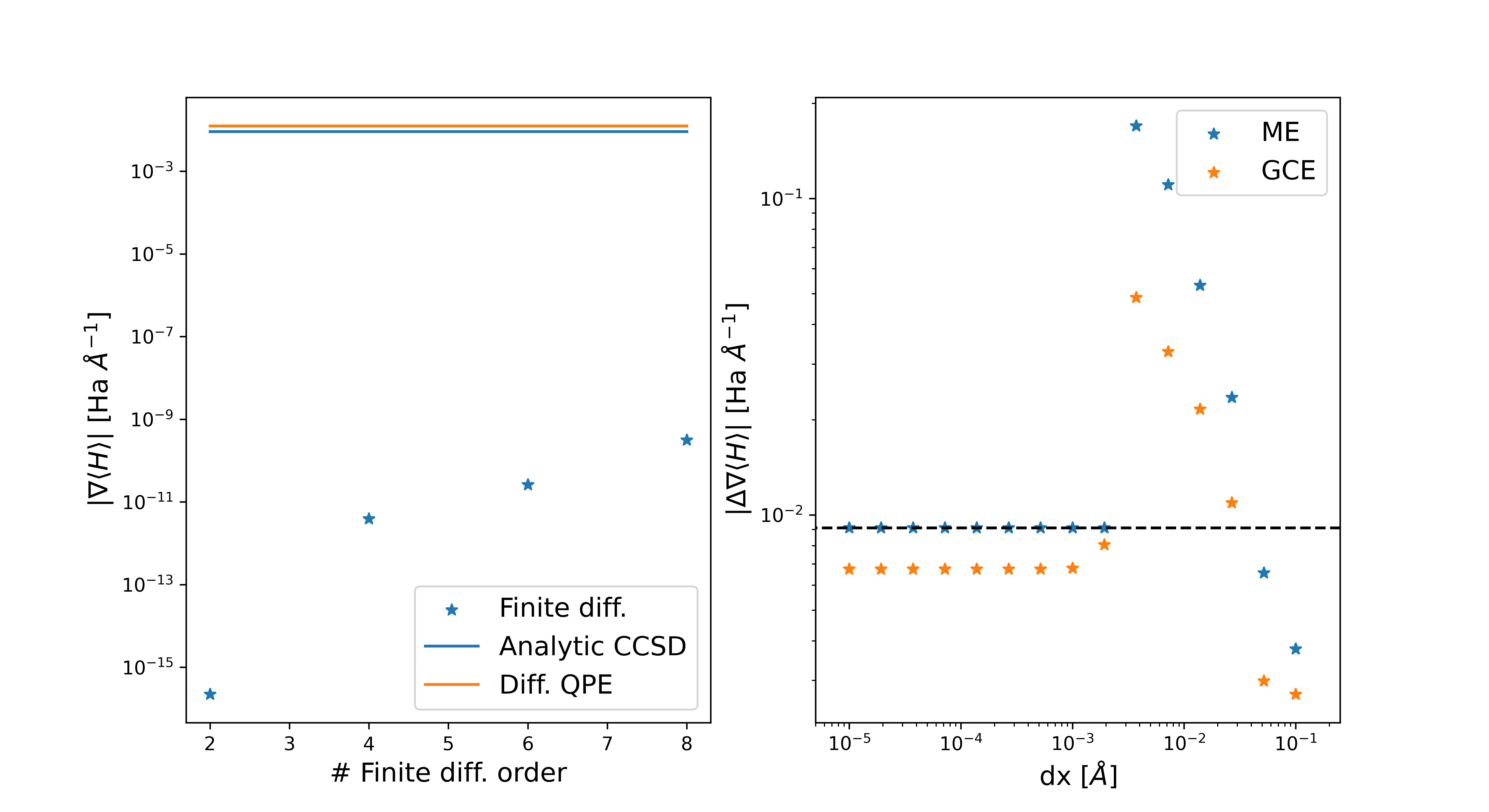}};

\node (a) at (-0.95, 2.0) {(a)};

\node (b) at (6.15, 2.0) {(b)};

\end{tikzpicture} \caption{Effects of discretization in the calculation of the gradients using finite difference methods. a) Norm of the energy gradient norm with respect to the nuclear positions for H$_3^{*}$ at the HF equilibrium geometry. The blue stars represent values obtained with a finite difference approximation to different orders using the majority rule estimator. The solid blue line depicts the analytically computed gradient at the CCSD level of theory, while the orange solid line corresponds to the gradient obtained with the fully differentiable pipeline described in Fig.\,\ref{computational_graph}. b) Gradient estimation error using a finite difference method (order 2) as a function of step size. The blue stars represent values obtained via the majority estimator, while the orange stars are obtained using the estimator developed in this work. All calculations were performed using $t$=13 readout qubits. The dashed line corresponds to the error when the gradient calculated with the QPE is zero.}
\label{finited_diff2}
\end{figure}

The last condition derives from the fact that for a signal to be detected with the finite difference approximation not only must the step size be small enough to avoid truncation errors, but it must also be such that function variations at different points can be spotted according to the QPE grid resolution. As shown in Fig.\,\ref{finited_diff2}, even high-order approximations to the gradient fail unless both conditions are satisfied.
In particular, Fig.\,\ref{finited_diff2}a shows the gradient norm $|\nabla \langle H \rangle|$ (used as sensitivity metric) computed with a central finite difference approximation at different orders (blue stars) adopting an ME. For comparison, we reported as a solid blue line the analytical result at the CCSD computational level (which is exact for a two-electron system as the one considered in this case) and as a solid orange line the gradient norm resulting from the differentiable QPE. As we can see, even using a degree 8 finite difference approximation with a step size of $\delta x = 10^{-3} \text{\AA}$ the resolution enabled by a $t=13$ grid does not allow us to detect any signal. In Fig.\,\ref{finited_diff2}b we compare the error on the gradient (reported as the difference from the CCSD value) estimated with a degree 2 finite difference method using the GCE (orange stars) and the ME (blue stars). The black dashed line is the exact gradient norm. Points lying on (above) this line are associated with null (or pointing in the wrong direction) gradients: in all these cases the inaccuracy makes the calculation useless. We see that the smooth GCE enables the computation of the gradient with very small step sizes, whereas the ME fails to detect any signal. Upon increasing $\delta x$ the estimated gradient starts suffering from truncation errors. Eventually, for $\delta x \approx 0.1 \AA$ the estimated gradients becomes more accurate. However, these step size values have already proven to be unreliable in standard quantum chemistry software implementations.

These results show that using finite differences with the ME may not be feasible on a small scale fault-tolerant computer (e.g., using $t \approx 10-20$), despite the good asymptotic scaling shown in Ref.\cite{o2022efficient}.

Summarizing the comparative analysis of this section, our approach currently suffers from a finite-shot sampling cost which makes it unfavourable compared to other strategies such as \ref{hellman-feynman} and \ref{qsvt-gradient} which saturate the Heisenberg limit. Nonetheless, our algorithm does not necessitate the preparation of the exact ground state on the quantum computer, a requirement that, on the other hand, is mandatory in both \ref{hellman-feynman} and \ref{qsvt-gradient}. This condition could potentially limit the application of these methodologies to regimes in which classical computation proves to be more efficient than quantum computation\cite{lee2023evaluating,fomichev2023initial}. Further, the proposed strategy is more flexible when it comes to considering extensions of this algorithm towards higher-order derivatives. In fact, the extension to higher order derivatives of the Hellmann-Feynman type of approach or the gradient-based estimation algorithm requires an exponentially higher\cite{o2019calculating} cost or redesigning the same algorithm, respectively. 

The goal of future work will be to restore optimal (Heisenberg) scaling of the approach proposed here by leveraging techniques such as compressed sensing\cite{yi2023quantum} or amplitude amplification \cite{brassard2002quantum} tailored to an automatically differentiable framework. Another essential direction of research to align this strategy with other proposals in the literature and possibly improve on them concerns reducing the cost of differentiation itself. Currently, quantum differentiation methods exhibit backpropagation-like scalings for only a specific class of circuits\cite{bowles2023backpropagation}.


\section{Geometry optimization}\label{results}

Here we report the numerical results of our implementation of the GCE described in Sec. \ref{GCE} coupled to the automatically differentiable pipeline provided by PennyLane. The latter, thanks to the differentiable quantum chemistry feature\cite{arrazola2021differentiable}, ensures the end-to-end differentiability of the procedure and enables its application to the calculation of energy gradients of molecular Hamiltonians, which is of utmost importance in the field of quantum chemistry. 

In addition to numerically validating the algorithm proposed in this study, the calculations themselves represent a novelty. To the best of our knowledge, only two other papers (focusing on fault-tolerant algorithms) in the literature report similar calculations: O'Brien \textit{et al.}\cite{o2019calculating} calculated the equilibrium bond length of H$_2$ with a minimal basis set on a real quantum processor, and, more recently, Sugisaki \textit{et al.}\cite{sugisaki2022quantum} developed a protocol to optimize molecular geometries using a Bayesian phase difference algorithm and applied it to different molecules of which the largest is N$_2$ with a Complete Active Space CI of 6 electrons in 6 orbitals (6e, 6o).

\subsection{Computational details}

All calculations were performed using a locally modified version of PennyLane\cite{bergholm2018pennylane} v0.24. The code generating all the data of this section can be found at \cite{qpe_code}.

We studied the trihydrogen cation and the formaldehyde. For the first molecule we used a gradient descent optimizer with stepsize of 0.01 $\AA$ as implemented in PennyLane, while for the second molecule we used an in-house automatically differentiable implementation of the BFGS optimizer \cite{nocedal1999numerical}.

Reference calculations at the Full Configuration Interaction (FCI) level were performed with the Psi4\cite{turney2012psi4} code. CASCI calculations on the CH$_2$O molecule were carried out with the PySCF code\cite{sun2018pyscf}. The STO-3G basis set was employed in all calculations. For the H$_3^+$ molecule, adopting this basis set entails using six qubits to encode the wave function into the system register by Jordan-Wigner encoding\cite{jordan1993algebraic}. For the CH$_2$O molecule, the active space contained two electrons in the four highest lying MOs. For all calculations we used $k=1000$ to smooth the boxcar function and a window width $h = \frac{8}{2^t}$. We set the temperature for the tempered-softmax distribution (Eq.\,\ref{tempered_softmax}) at $T = 0.0035$. We found that higher values of $k$ (i.e., $k > 5000$) in some cases lead to numerical instabilities. We recall that a thorough analysis of the window size effects is given in Sec.\,\ref{GCE} (Fig.\,\ref{window_size_eff}). We set the number of readout qubits to $t = 13$ for the H$_3^+$ molecule and $t = 11$ for the CH$_2$O molecule. Overall we exploited 19 qubits in both cases.

It is worth noting that QPE methods always involve a rescaling of the Hamiltonian to avoid aliasing of the energy found. In practice, it means working with a rescaled Hamiltonian $H_{SC} = \frac{H_{mol} - E_{min}\mathcal{I}}{\Delta E}$, where $\Delta E$ is the relevant spectral range. However, this procedure increases the computational cost, which scales as $\mathcal{O}(\Delta E \epsilon^{-1}$) rather than $\mathcal{O}(\epsilon^{-1})$. In all our numerical simulations we mitigated this issue by working with purely electronic Hamiltonians (i.e., adding \textit{a posteriori} the frozen core contribution and/or the internuclear interaction terms).

\subsection{Trihydrogen cation (H$_3^+$)}

As a first application of our differentiable pipeline, we show two examples of geometry optimization of H$_3^+$. This system, being the simplest polyatomic molecule, has been extensively studied for benchmarking \textit{ab initio} models\cite{tennyson1995spectroscopy, polyansky2012spectroscopy} and it has also attracted attention because it is one of the most common species in molecular hydrogen plasmas found in  interstellar clouds and planetary ionospheres\cite{mccall2000h3+}. 

\begin{figure}[h!]

\begin{tikzpicture}[node distance=cm,
    every node/.style={fill=white, font=\sffamily}]
    
    \node (figure) at (0,0) 
    {\centering
    \includegraphics[width = \textwidth]{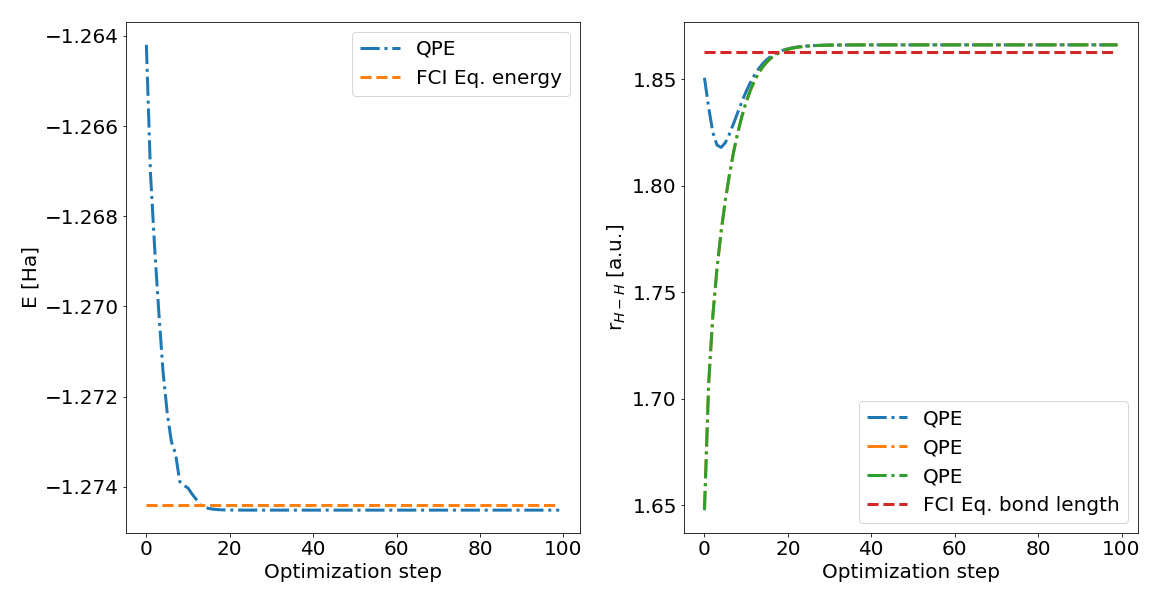}};

\node (a) at (-0.5, 2.45) {(a)};

\node (b) at (7.25, 2.45) {(b)};

\end{tikzpicture}
    \caption{Ground state geometry optimization of the H$_3^+$ molecule. a) Convergence of the ground-state energy E and b) bond length $r_{H-H}$. Dash-dot lines represent values obtained during the optimization at the QPE level, dashed lines are reference values obtained at the FCI level.}
\label{h3_gs}
\end{figure}

In Fig.\,\ref{h3_gs} we report a convergence plot of the geometry optimization for the ground state which is stable if the hydrogen atoms are arranged in a triangular shape\cite{aguado2000global}. The initial state used for this calculation is the Hartree-Fock (HF) state, which after Jordan-Wigner mapping reads $|HF\rangle = |110000\rangle$. The initial geometry was set equal to an isosceles triangle with bond lengths slightly distorted from the HF equilibrium geometry. Note that in Fig.\ref{h3_gs}b the dash-dot lines refer to different bond lengths and that the green and yellow lines are superimposed. As we can see, after 20 iterations the optimization is able to recover the correct equilateral shape ($D_{3h}$ symmetry) in agreement, within chemical accuracy, with the FCI optimized geometry. Interestingly, we notice that the energy obtained using the QPE is slightly below the FCI reference value (orange dashed line in Fig.\ref{h3_gs}a) due to the fact that the QPE is not a variational method.

\begin{figure}[h!]

\begin{tikzpicture}[node distance=cm,
    every node/.style={fill=white, font=\sffamily}]
    
    \node (figure) at (0,0) 
    {\centering
    \includegraphics[width = \textwidth]{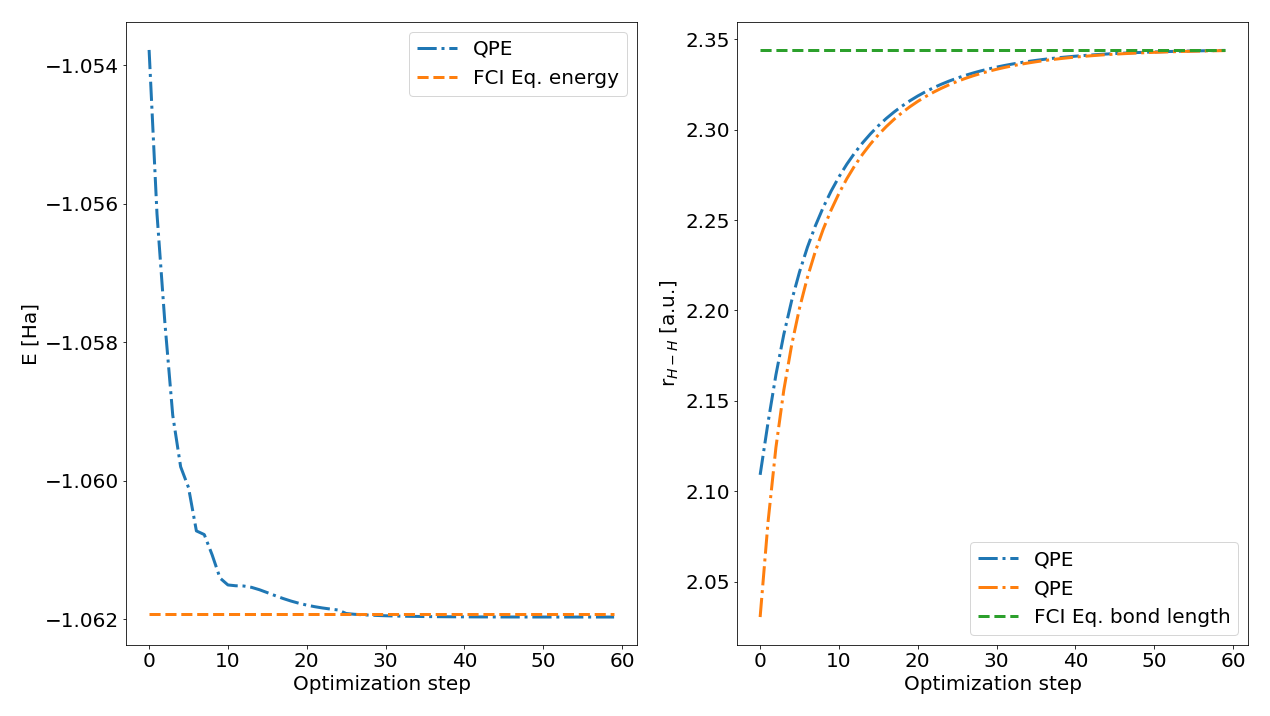}};

\node (a) at (-0.5, 2.75) {(a)};

\node (b) at (7.25, 2.75) {(b)};

\end{tikzpicture}
    \caption{Geometry optimization of the H$_3^+$ molecule in its first excited triplet state ($|\text{T}_0\rangle$). a) Convergence of the ground-state energy E and b) bond length $r_{H-H}$. The dash-dot lines represent values obtained during the optimization at the QPE level, while the dashed lines are reference values obtained at the FCI level.}
\label{h3_trip}
\end{figure}

As a second application example, we consider the first excited triplet state of the same molecule. Now the initial state is a spin-flipped doubly excited determinant built upon the HF configuration, i.e., $|\psi\rangle = t_{01}^{35}|HF\rangle = |0001010\rangle$. As  reported in the literature\cite{sanz2001lowest}, the triangular geometry is not stable in this excitation and undergoes photodissociation. On the other hand, a linear structure is a stable configuration for the $|T_0\rangle$ state; thus, we initialized the geometry as linear with the two terminal hydrogens at different distances from the central one (see Fig.\,\ref{h3_trip}b). Again, we see that the optimization allows us to reach energy and bond length with errors below chemical accuracy (i.e., $\Delta E < 1\, \text{m}Ha$ and $\Delta r < 0.01\, \AA$).

In the next section we showcase an application to a larger molecule to understand the impact of our procedure when considering a more elaborate electronic structure.

\subsection{Formaldeyhde (CH$_2$O)}

Here we show two examples of geometry optimization carried out with our automatically differentiable QPE on a larger molecular system. The formaldehyde molecule is studied in its ground state and first triplet state ($|T_0\rangle$) which corresponds to the spin-flip transition $^3(n \rightarrow \pi^*)$. These systems have been extensively studied both theoretically and experimentally\cite{krylov2000excited,angeli2005casscf,clouthier1983spectroscopy} as representative systems of carbonyl-chemistry and, more broadly, because of their role in photochemistry.

Throughout the optimization we set the $|HF\rangle$ state as the input state of the QPE. For the active space considered, it is $|HF\rangle = |11000000\rangle$. In Tab.\,\ref{form_gs} we report the results of our optimization and compare them with a geometry optimization performed at the CASCI level (using the same active space) and with experimental data from \cite{tables1985mw} as reported in \cite{schuetz2003geometry}.

As we can see the predicted length of the CO double bond is within chemical accuracy ($\Delta r < 0.01\, \AA$) considering both the theoretical calculations at the CASCI level and the experimental values as reference. Also, considering the predicted angles, an error of $\Delta \theta \approx 2^{\circ}$ represents a good agreement between theory (both methods) and experiment. Instead, by looking at the predicted CH bond length we can see that there is a more significant discrepancy between the theoretical methods ($\Delta r \approx 0.05\, \AA$). Such a difference may be ascribed to different factors: (i) bias ensuing from the external parameters $T$ and $k$, (ii) possible numerical instabilities encountered in the AD pipeline and (iii) differences in the optimization procedure and the underlying Hartree-Fock solver. Noteworthy, the results obtained using the differentiable QPE approach agree better with the experimental values than the other theoretical results. The agreement with the experimental results in Ref.\,\cite{takagi1963millimeter} is even within the limits of chemical accuracy. This fact can be attributed to error cancellation as the active space is truncated.

\begin{table*}[h!]
    \small
    \centering
    \caption{Optimized geometry of formaldehyde in the $|\text{GS}\rangle$ electronic state. Distances are expressed in $\text{\AA}$, angles in degrees. Experimental errors are reported in brackets.}
    \label{form_gs}
    \begin{tabular*}{\textwidth}{@{\extracolsep{\fill}}lccc}
        \hline
        \hline
        & CASCI(2e, 4o) & QPE(2e, 4o) & Exp.\cite{tables1985mw} \\
        r$_{\text{CO}}$ & 1.212 & 1.211 & 1.21(1)\\ 
        r$_{\text{CH}}$ &  1.055 & 1.106 &  1.12(1)\\
        $\angle_{OCH}$ &  123.18 & 122.96 & 121(1)\\
        $\angle_{OCH}$ &  123.18 & 122.96 & 121(1)\\
        $\angle_{HCH}$ &  113.64 & 114.08 & - \\
    \end{tabular*}
    \label{numerical_details}
\end{table*}

We now turn to the discussion of the optimization of the molecule in its first triplet state ($|T_o\rangle$). Here, all along the optimization procedure, we set as initial state the triplet state built upon excitation of the HF determinant, $|\psi\rangle = t_0^2|HF\rangle + t_1^3|HF\rangle$ where $t_i^j$ is a single excitation from spin-orbital $i$ to $j$. We recall that in our convention for labeling the orbitals, $\alpha/\beta$ spins are interleaved and that spin-orbital labeling starts with the leftmost qubit being indexed with 0.

As we can see from Tab.\,\ref{form_trip}, bond lengths and angles from QPE are within chemical accuracy of the CASCI reference values, thus further supporting that the GCE estimator coupled with the automatic differentiation pipeline of PennyLane is able to reproduce both total energies and forces. However, with respect to the experiment (and regardless of the theoretical method employed), the chosen active space causes an error of $\approx 0.02 \AA$, leading to overestimation of the CO bond length and underestimation of the CH bond length. This is in agreement with other results reported in  the literature for calculations in restricted active spaces. For example, Angeli et al.\cite{angeli2005casscf} report errors of $0.062$ and $0.028\, \text{\AA}$ for the CO bond length and of $0.005$ and $0.028\, \text{\AA}$ for the CH bond using two different active spaces and including orbital optimization.

\begin{table*}[h!]
    \small
    \centering
    \caption{Optimized geometry of formaldehyde in its first triplet electronic state ($|\text{T}_0\rangle$). Distances are expressed in $\text{\AA}$, angles in degrees.}
    \label{form_trip}
    \begin{tabular*}{\textwidth}{@{\extracolsep{\fill}}lccc}
        \hline
        \hline
        & CASCI(2e, 4o) & QPE(2e, 4o) & Exp.\cite{clouthier1983spectroscopy} \\
        r$_{\text{CO}}$ & 1.320 & 1.325 & 1.307\\ 
        r$_{\text{CH}}$ &  1.063 & 1.060 & 1.084 \\
        $\angle_{OCH}$ &  117.36 & 119.75 & - \\
        $\angle_{OCH}$ &  116.72 & 119.74 & - \\
        $\angle_{HCH}$ &  125.92 & 120.51 & 121.8\\
    \end{tabular*}
    \label{numerical_details}
\end{table*}

The above results demonstrate not only the effective application of the automatically differentiable QPE to a non trivial chemical system but also emphasize the importance of combined numerical and cost analysis when proposing an algorithm. Indeed, our results highlight the need for specific-domain knowledge and stress the importance of gaining practical experience, as similarly demonstrated in the literature for standard computational quantum chemistry, to obtain results in line with other methods and experiments. In particular, our analysis indicates that the extension of experiments to the optimization of high-energy excited states requires on-the-fly assessments to understand how the displacement of atoms induces changes in the electronic structure and accordingly determine how to modify the input molecular state in the quantum circuit. As a possible future development, we will consider the design of a cheap and fast algorithm to make this evaluation feasible.

\section{Conclusions}\label{conclusions}

We showed that a smooth estimator allows the seamless integration of the QPE algorithm within the framework of automatic differentiation. Initially proposed in Ref.\cite{cruz2020optimizing}, the circular estimator was confined to the case when exact eigenstates of an Hamiltonian are known and can be efficiently prepared on a quantum register. Here we overcomed this fundamental issue unlocking its practical use.
Along with a new, more accurate estimator for the QPE and the possibility of integrating the algorithm in a new programming environment, here we demonstrate a new off-the-shelf strategy to devise quantum algorithms enabling the simulation of molecular properties. 

To this end, we showcased the calculation of optimized geometries outperforming the existing literature both from the point of view of system size (number of atoms) and target states, as we present calculations beyond the singlet ground state geometry optimization. The latter feature is of utmost importance when it comes to the simulation of photochemical experiments.
The next, natural step of this work is the application of this strategy to the simulation of spectroscopy experiments.


Despite the potential of the approach reported in this study, some aspects currently limit its scalability to larger systems. In particular, we see that the reconstruction of the measurement distribution requires a substantial computational overhead, which can be mitigated by improving the wave function guess. Further, a straightforward improvement may result from the inclusion of a cosine or Kaiser window in the phase estimation circuit. This concentrates the information in the main lobe of the distribution and may lower the number of relevant bitstrings to sample for the circular average. This solution has already proven to give a cubic speedup for the majority rule estimator\cite{rendon2022effects} and may provide similar speedups for the GCE. Alternatives strategies to circumvent this issues could resort to recent variations of the quantum phase estimation such as in Refs.\cite{ni2023low, yi2023quantum, ding2023even}. 


It should also be emphasized that when developing a quantum algorithm we are not only concerned with finding a solution to a useful problem but also with ensuring that the runtime of the algorithm is feasible and advantageous compared to other solutions, including classical ones. Despite the favorable scaling of QPE, the answer to this question in the field of quantum chemistry is still open\cite{delgado2022simulating, zini2023quantum, lee2023evaluating, rubin2023fault}. For what concerns the algorithm proposed in this work, further research is necessary to arrive at cost-effective differentiable solutions. Our study further underlines the importance of expanding the classes of circuits that allow for backpropagation scaling in derivative calculations\cite{bowles2023backpropagation}. 

\section*{ACKNOWLEDGMENTS}

The authors acknowledge the University of Padova Strategic Research Infrastructure Grant 2017: “CAPRI: Calcolo ad Alte Prestazioni per la Ricerca e l’Innovazione” for HPC usage. D.C. is grateful to MIUR ”Dipartimenti di Eccellenza” under the project  Nanochemistry  for  energy  and Health (NExuS) for funding the PhD grant. D.C. and S.C. also acknowledge fundings from the MUR PNRR-PRIN2022 (PNRR M4C2 Inv 1.1) grant 2022W9W423 "Quantum computing for Computational Chemistry and Materials Science" funded by the European Union - NextGenerationEU.

\appendix

\section{The physical meaning of differentiating through the measurement distribution}
\label{physical_meaning}

In this section we derive analytic expressions for the energy derivative as defined by Eq. \ref{def_generalized_circular_est}. These expressions allow us to understand the physical origin of the automatically computed gradient.

The main idea is that we can recast the whole differentiable QPE procedure in this expression:

\begin{equation}
    E \approx \mathcal{S} \big [ \int \langle \Psi | e^{-i H t} | \Psi \rangle e^{i \omega t} dt \big ] = \mathcal{S} \big [ I(\omega) \big ]
\end{equation}
where $\mathcal{S}$ is a differentiable functional corresponding to the smoothing procedure outlined in Fig.\,\ref{scheme_distributions} of the main text.

Accordingly, we can write explicitly the energy derivative differentiating the last expression:

\begin{equation}
    \frac{\partial \mu}{ \partial \textbf{R}} = \frac{\delta \mathcal{S}}{\delta I} \frac{\partial I}{\partial \textbf{R}}
\end{equation}

Here, our focus is on the second factor $\frac{\partial I}{\partial \textbf{R}}$, as the first does not provide any information about the physics of the system but concerns the dependence of our estimator on the parameters denoted as $T$ and $k$ in the main text.

By expanding $|\Psi\rangle$ on the Hamiltonian eigenbasis we get:

\begin{equation}
    I(\omega, \textbf{R}) = \sum_i |c_i|^2(\textbf{R}) \delta(\phi_i(\textbf{R}) - \omega)
\end{equation}
where $\phi_i = f(E_i)$. As expressed in the introduction (Sec.\,\ref{intro_to_QPE}).

Differentiating and accounting for the finite precision of the quantum register grid we obtain: 

\begin{equation}\label{distribution_der}
    \frac{\partial I}{\partial \textbf{R}} = \sum_i \frac{\partial |c_i|^2}{\partial \textbf{R}} \frac{\partial P}{\partial \phi_i} \frac{\partial \phi_i}{\partial \textbf{R}} 
\end{equation}
Where $P$ is the parent distribution kernel defined in Eq.\ref{probabilities_sampling} and $i$ is a label running over the electronic eigenstates.

In our approximation, i.e. Eq. \ref{decoupling_approximation}, the last equation reduces to:

\begin{equation}\label{distribution_der}
    \frac{\partial I}{\partial \textbf{R}} \approx \frac{\partial |c_i|^2}{\partial \textbf{R}} \frac{\partial P}{\partial \phi_i} \frac{\partial \phi_i}{\partial \textbf{R}} 
\end{equation}
where $|i\rangle$ is the only eigenstate that contributes to the probability $P$.

We again focus on the terms that bring the molecular physics into the algorithm. Therefore we do not comment the term $\frac{\partial P}{\partial \phi_i}$ which can be obtained analytically from Eq.\,\ref{probabilities_sampling}.

By applying the chain rule to the eigenphase derivative we can easily see the emergence of the true molecular energy derivatives: $\frac{\partial \phi_i}{\partial \textbf{R}} = \frac{\partial \phi_i}{\partial f}\frac{\partial f}{\partial E_i}\frac{d E_i}{d \textbf{R}}$. The dependence on $f$ is a purely algorithmic choice and will not be considered here. Instead, following Ref.\,\cite{o2022efficient} we directly relate the force calculated with respect to the i-th eigenstate to molecular integrals derivatives as:

\begin{equation}\label{energy_der}
    \frac{d E_i}{d \textbf{R}} = \Bigg (\sum_{pq} \gamma_{pq} \frac{d h_{pq}}{d \textbf{R}} + \sum_{pqrs} \Gamma_{pqrs} \frac{d g_{pqrs}}{d \textbf{R}}\Bigg ) - \Bigg ( \sum_{pqm} \gamma_{pq} h_{mq} \frac{d S_{mp}}{d \textbf{R}} + 2 \sum_{pqrst} \Gamma_{pqrs} g_{tqrs} \frac{d S_{tp}}{d \textbf{R}} \Bigg )
\end{equation}
where $\gamma_{pq}$ and $\Gamma_{pqrs}$ are the one- and two-electrons reduced density matrices\cite{helgaker2013molecular} for the i-th eigenstate respectively, $h$ and $g$ are the one- and two-electrons integrals representing all the interactions in the molecular Hamiltonian (i.e., electron-nuclear interactions and electron-electron interactions). All indexes of the summation run over the spin-orbitals. $S$ is the basis overlap matrix.

To obtain a complete expression for the energy derivative, we evaluate $\frac{\partial |c_u|^2}{\partial \textbf{R}}$. We consider a real-valued wavefunction (without any loss of generality as long as we are not neglecting the presence of external magnetic fields) for which we can write the first-order response of the wavefunction with respect to a perturbation (in our case the perturbation is a nuclear displacement) as\,\cite{helgaker1988analytical}:

\begin{equation}
\label{coeff_der}
    \frac{\partial c_\mu}{\partial \textbf{R}} = - \sum_{\nu} \mathcal{G}_{\mu\nu}^{-1} \frac{\partial \mathcal{F}_\nu}{\partial \textbf{R}}
\end{equation}
where $\mathcal{G}$ and $\mathcal{F}$ are the electronic Hessian and the electronic gradient, respectively, defined as:

\begin{equation}
    \mathcal{G}_{\mu\nu} = \frac{\partial^2 E}{\partial c_{\mu} \partial c_{\nu}} = 2 (\langle \mu | H | \nu \rangle -  E_{GS} \langle \mu | \nu \rangle) 
\end{equation}

\begin{equation}\label{electron_grad}
    \mathcal{F}_{\mu} = \frac{\partial E}{\partial c_{\mu}} = 2 \langle \mu |PH|\Psi\rangle 
\end{equation}
The previous equations hold for the derivatives of the ground state coefficients. Moreover, we point out that $|\mu\rangle \, \, \text{and} \, \, |\nu\rangle$ are $N$-electrons Slater determinants. The expressions in the Hamiltonian eigenbasis can be recovered considering that the two sets of coefficients are related by the basis change $\textbf{C}_{\{|u\rangle\}} = \textbf{V}^{\dagger}\textbf{C}_{\{|\mu\rangle\}}\textbf{V}$ (where $\textbf{V}$ is the matrix that diagonalizes the Hamiltonian). Finally, in Eq.\,\ref{electron_grad} we have introduced the projection operator $P = \mathbf{1} - |\Psi\rangle\langle\Psi|$.

Using Eq. \ref{coeff_der} and $\frac{\partial |c_u|^2}{\partial \textbf{R}} = 2 c_u \frac{\partial c_u}{\partial \textbf{R}}$, we can attribute a precise physical meaning to the overlap derivative term: in analogy with \textit{Hooke's law}, the wavefunction relaxation is proportional to the ``force" $-\frac{\partial \mathcal{F}_\nu}{\partial \textbf{R}}$ and inversely proportional to the ``force constant" $\mathcal{G}$. Indeed, in the absence of external perturbation (i.e. nuclear displacement), the electronic gradient is null and the wavefunction is stable. 

Finally, adding the contribution of Eq.\,\ref{energy_der}, we can reconstruct the meaning of Eq. \ref{distribution_der}. 
In essence, Eq.\,\ref{distribution_der} tells us that the measurement distribution induced by the QPE circuit varies as nuclei are displaced due to the variation of the electronic energy (Eq.\,\ref{energy_der}) and the wavefunction coefficients (Eq.\,\ref{coeff_der}). These contributions are singled out and modulated by the functional $\mathcal{S}$, other algorithmic choices (e.g., $f \propto \text{cos}^{-1}$ when employing a qubitization\cite{low2019hamiltonian} based QPE), and the number of readout qubits entering the quantity $\frac{\partial P}{\partial \phi_i}$.

\section{Effect of sampling noise}
\label{num_sampling_noise}

\begin{figure}[h!]
\begin{tikzpicture}[node distance=cm,
    every node/.style={fill=white, font=\sffamily}]
    
    \node (figure) at (0,0) 
    {\centering
    \includegraphics[width = \textwidth]{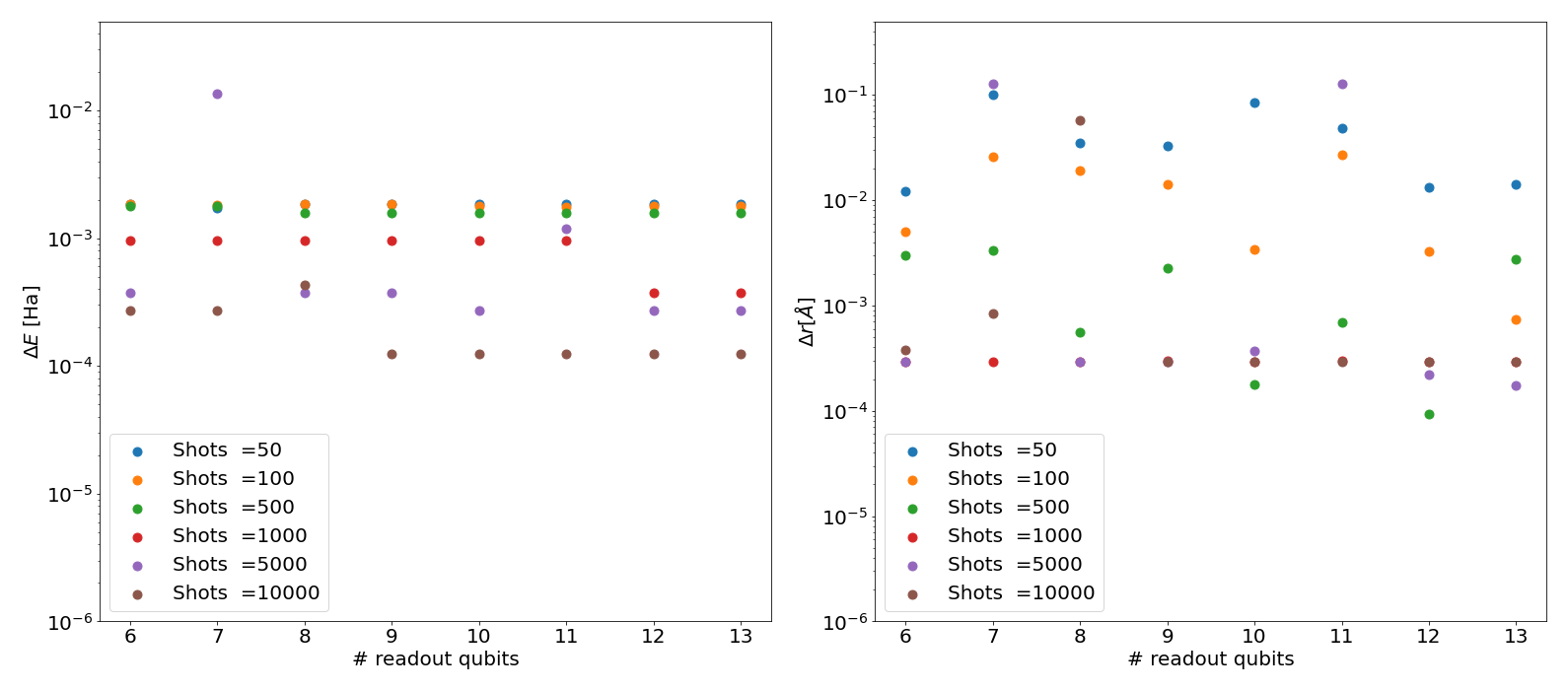}};

\node (a) at (-0.75, 2.75) {(a)};

\node (b) at (7.25, 2.75) {(b)};

\end{tikzpicture}
\caption{Ground state geometry optimization under sampling noise. (a) Energy error ($\Delta E$) and bond length error (b) ($\Delta r$) with respect to the FCI optimized solution as a function of number of readout qubits varying the number of collected samples for reconstructing the parent distribution (shown with different colors). We remark that each point is an independent optimization run.}
\label{more_on_sampling_noise}
\end{figure}

In Fig.\,\ref{more_on_sampling_noise}, we illustrate the effect of sampling on the geometry optimization of H$_3^+$ in its singlet ground state using the strategy defined in the main text, see Sec.\,\ref{intro_idea} and Sec.\,\ref{GCE}.

Both in terms of energy error ($\Delta E$, Fig.\,\ref{more_on_sampling_noise}a) and bond length error ($\Delta r$, Fig.\,\ref{more_on_sampling_noise}b), the estimate of the number of samples to be measured outlined in Fig.\,\ref{optimal_window_size} is consistent with the findings reported here. Indeed, the initial state used to perform the geometry optimization is always the Hartree-Fock determinant, which, starting from the initial geometry (the same one used for the calculations presented in the main text), exhibits an overlap with the FCI solution greater than 90\%.

Interestingly, by looking at Fig.\,\ref{more_on_sampling_noise}, we may be tempted to conclude that the generalized circular estimator variance does not actually depend on the number of readout qubits, as very similar trends are observed with respect to the number of shots regardless of the number of readout qubits; also, the final errors with respect to the exact geometry look quite independent on $t$. As we have discussed in Sec.\ref{gce_formal_analysis}, this is not the case. The results shown in Fig.\,\ref{more_on_sampling_noise} arise from the interplay of two effects: (i) the window sizes that we have used in these calculations shrink exponentially with the number of qubits, as reported in Fig.\,\ref{window_size}. This implies that the number of bitstrings included in the circular average (for the calculations reported in Fig.\,\ref{more_on_sampling_noise}) is always the same regardless of the number of readout qubits. Clearly, this approach works also with a lower number of qubits because the input state used allows sampling from a large number of bitstrings without adding information from spurious excited states. (ii) These results do not refer to the a single-point energy evaluation but are obtained from different independent optimization processes. Therefore the errors reported here combine the errors on the function \- and gradient \- evaluations with those in the optimization process.

\bibliography{bibliography}

\end{document}